\documentclass[reprint,superscriptaddress,amsmath,amssymb,aps,prb]{revtex4-2}

\usepackage{graphicx,dcolumn,bm,xcolor,soul}
\usepackage[colorlinks,citecolor=blue,urlcolor=blue,linkcolor=blue]{hyperref}

\definecolor{airforceblue}{rgb}{0.36, 0.54, 0.66}
\definecolor{antiquefuchsia}{rgb}{0.57, 0.36, 0.51}
\definecolor{asparagus}{rgb}{0.53, 0.66, 0.42}
\definecolor{beaver}{rgb}{0.62, 0.51, 0.44}
\definecolor{bole}{rgb}{0.47, 0.27, 0.23}
\definecolor{cadet}{rgb}{0.33, 0.41, 0.47}
\definecolor{camouflagegreen}{rgb}{0.47, 0.53, 0.42}
\definecolor{charcoal}{rgb}{0.21, 0.27, 0.31}
\definecolor{darkcerulean}{rgb}{0.03, 0.27, 0.49}
\definecolor{darkelectricblue}{rgb}{0.33, 0.41, 0.47}
\definecolor{darkolivegreen}{rgb}{0.33, 0.42, 0.18}
\definecolor{deepchestnut}{rgb}{0.73, 0.31, 0.28}
\definecolor{feldgrau}{rgb}{0.3, 0.36, 0.33}
\definecolor{frenchlilac}{rgb}{0.53, 0.38, 0.56}
\definecolor{glaucous}{rgb}{0.38, 0.51, 0.71}
\definecolor{lapislazuli}{rgb}{0.15, 0.38, 0.61}
\definecolor{brickred}{rgb}{0.8, 0.25, 0.33}
\definecolor{celestialblue}{rgb}{0.29, 0.59, 0.82}
\definecolor{darkpastelblue}{rgb}{0.47, 0.62, 0.8}

\colorlet{RED}{beaver!80}
\colorlet{BLUE}{airforceblue}
\colorlet{BLACK}{charcoal!90}
\definecolor{GRAY}{RGB}{245,245,245}

\usepackage{pgfplots}
\usepgfplotslibrary{groupplots}
\pgfplotsset{compat=1.18}
\usetikzlibrary{patterns.meta}
\pgfplotsset{
  every axis/.style={
    tick align=inside,
    tick pos=both,
    enlargelimits=false,
    font=\normalsize,
    xlabel style={font=\normalsize},
    ylabel style={font=\normalsize},
    tick label style={font=\normalsize},
    legend style={font=\normalsize, cells={anchor=west}},
    line width=1pt,
    mark size=2.5pt,
  }
}
\usetikzlibrary{calc,positioning,fit,shapes.geometric,arrows.meta}

\colorlet{CtaskL}{darkolivegreen!30}
\colorlet{CtaskS}{blue!15!white}
\colorlet{CGPU}{red!15}
\colorlet{CThr}{blue!15}
\colorlet{GPUred}{RED}
\colorlet{ThrBlue}{BLUE}
\colorlet{MPIsum}{gray!80!black}

\tikzset{
  base/.style = {align=center, inner sep=3pt, outer sep=1pt, rounded corners=2pt},
  box/.style = {base, draw=black, line width=1pt, fill=white, minimum width=22mm, minimum height=10mm},
  head/.style = {box}, 
  image/.style = {box}, 
  pool/.style ={box},
  rank/.style ={box},
  taskL/.style     ={draw=black, line width=0.9pt, fill=CtaskL, minimum height=10mm, minimum width=27mm, rounded corners=1.5pt},
  taskS/.style     ={draw=black, line width=0.9pt, fill=CtaskS, minimum height=10mm, minimum width=27mm, rounded corners=1.5pt},
  gpu/.style       ={draw=darkolivegreen!30, line width=1.2pt, rounded corners=2pt, minimum height=15mm, minimum width=27mm, text=darkolivegreen},
  thr/.style = {draw=darkelectricblue!30, line width=1.2pt, rounded corners=2pt, minimum height=15mm, minimum width=27mm, text=darkelectricblue},
  flow/.style = {-, thick},
  groupbox/.style = {draw, rounded corners=3pt, line width=0.9pt, inner sep=3pt, fill=none},
  mpisum/.style = {<->, >=latex, draw=MPIsum, line width=1pt, shorten >=4pt, shorten <=4pt}, 
}

\usepackage[newfloat]{minted}
\usepackage[T1]{fontenc}
\usepackage{lmodern}

\newcommand{\eph}{e-ph}
\newcommand{\epw}[1]{EPW{#1}}

\newcommand{\vscf}{V^{\rm SCF}}
\newcommand{\pd}{\partial}
\newcommand{\kk}{\mathbf{k}}
\newcommand{\qq}{\mathbf{q}}
\newcommand{\RRp}{\mathbf{R}_{\rm p}}

\newcommand{\RRe}{\mathbf{R}_{\rm e}}
\newcommand{\ka}{\kappa\alpha}

\newcommand{\rr}{\mathbf{r}}
\newcommand{\Na}{N_{\rm a}}
\newcommand{\Nb}{N_{\rm b}}
\newcommand{\Nm}{N_{\rm m}}
\newcommand{\Nkk}{N_{\kk}}
\newcommand{\Nqq}{N_{\qq}}
\newcommand{\NRRp}{N_{\rm p}}
\newcommand{\NRRe}{N_{\rm e}}

\newcommand{\Npo}{N_\text{pool}}
\newcommand{\Nim}{N_\text{image}}
\newcommand{\Nth}{N_\text{thread}}
\newcommand{\Teph}{N}

\newcommand{\kkp}{\kk'}
\newcommand{\qqp}{\qq'}
\newcommand{\Nkkp}{N_{\kk'}}
\newcommand{\Nqqp}{N_{\qq'}}

\begin{document}

\title{Electron-phonon physics at the exascale:\\[2pt]
A hybrid MPI-GPU-OpenMP framework for scalable Wannier interpolation} 

\author{Tae Yun Kim}
\affiliation{Oden Institute for Computational Engineering and Sciences, The University of Texas at Austin, Austin, Texas 78712, USA}
\affiliation{Department of Physics, The University of Texas at Austin, Austin, Texas 78712, USA}

\author{Zhe Liu}
\affiliation{Department of Physics, Applied Physics, and Astronomy, Binghamton University, Binghamton, New York 13902, USA}

\author{Sabyasachi Tiwari}
\affiliation{Oden Institute for Computational Engineering and Sciences, The University of Texas at Austin, Austin, Texas 78712, USA}
\affiliation{Department of Physics, The University of Texas at Austin, Austin, Texas 78712, USA}

\author{Elena R. Margine}
\affiliation{Department of Physics, Applied Physics, and Astronomy, Binghamton University, Binghamton, New York 13902, USA}

\author{Feliciano Giustino}
\email{fgiustino@oden.utexas.edu}
\affiliation{Oden Institute for Computational Engineering and Sciences, The University of Texas at Austin, Austin, Texas 78712, USA}
\affiliation{Department of Physics, The University of Texas at Austin, Austin, Texas 78712, USA}

\date{\today}

\maketitle

\noindent\textbf{ABSTRACT}

\smallskip

\noindent
We demonstrate a highly efficient GPU implementation of the Wannier interpolation of electron-phonon matrix elements in the \epw{} code. Building on a systematic analysis of the computational complexity of the algorithm for electron-phonon interpolation, we designed a GPU porting strategy that integrates naturally into the current \epw{} implementation, and is seamlessly portable to NVIDIA, AMD, and Intel GPUs. {We demonstrate this development via extensive benchmarks on conventional semiconductors such as silicon and monolayer MoS$_2$, as well as a large-scale application to topological stanene nanoribbons of width as large as 20\,nm, which was intractable with previous implementations. Compared to the single MPI parallelization scheme of EPW v5.9,} the resulting hybrid MPI-GPU-OpenMP scheme achieves up to 29-fold speedup on leadership-class supercomputers equipped with NVIDIA and Intel accelerators, namely Vista at the Texas Advanced Computing Center, Perlmutter at the National Energy Research Scientific Computing Center, and Aurora at the Argonne Leadership Computing Facility. This framework also achieves nearly ideal scalability up to thousands of GPU nodes on the Aurora supercomputer.  With this development, \epw{} is ready to support {electron-phonon physics calculations on exascale platforms.}

\bigskip

\noindent\textbf{INTRODUCTION}

\smallskip

The interaction between electrons and phonons is ubiquitous in condensed matter physics. Many physical properties of materials arise from the electron-phonon (\eph{}) interaction~\cite{Ziman1960,GiustinoRMP2017}: for example, the temperature dependence of the electrical conductivity, the absorption of light in indirect band gap semiconductors, the formation of localized polarons, and phonon-mediated superconducting pairing are all phenomena linked to the coupling between electrons and phonons~\cite{Mahan2000}.

In order to correctly capture \eph{} couplings throughout the Brillouin zone, it is necessary to employ specialized interpolation schemes since calculations of phonons using { density functional perturbation theory (DFPT)}~\cite{BaroniPRL1987,SavrasovPRL1992,GonzePRB1997,BaroniRMP2001} on fine wavevector grids are costly. One of the most popular strategies is \eph{} Wannier interpolation~\cite{GiustinoPRB2007}, which leverages the spatial localization of maximally-localized Wannier functions~\cite{Marzari1997,SouzaPRB2001,Marzari2012,Marrazzo2024}. This approach was initially demonstrated with the \epw{} code, and is now used by many related software packages~\cite{Zhou2021,Protik2022,Cepellotti2022,Marini2024}.

{
The field of \eph{} theory has seen important methodological advances in recent years.
Significant progress has been made in treating long-range electron-phonon interactions and dynamical quadrupoles~\cite{Antonius2015, Brunin2020} and density-matrix perturbation theory~\cite{Yang2022}.
Non-perturbative methods have emerged for systems with strong vibronic coupling~\cite{Zacharias2016, Monserrat2018, Kundu2023}.
These methodological developments are complementary to the Wannier interpolation method employed in the \epw{} code.
}

\epw{} is a Fortran/MPI program that specializes in calculations of \eph{} interactions and related materials properties~\cite{LeeNPJ2023,PonceCPC2016,Noffsinger2010}.
The program uses density functional theory (DFT)~\cite{HohenbergPR1964,KohnPR1965} and DFPT to compute the \eph{} matrix elements from first principles.
Even though \epw{} significantly reduces the cost of computing \eph{} matrix elements as compared to direct DFPT calculations, the ultra-fine Brillouin zone grids required to achieve predictive accuracy can still make workflows very demanding, even on thousands of nodes in modern supercomputers.

Since recently deployed exascale supercomputers rely heavily on graphics processing unit (GPU) accelerators~\cite{Top500project}, achieving optimal GPU performance has become a top priority in high-performance computing (HPC) applications.
{
GPU acceleration has become essential across many-body electronic structure codes.
Notable implementations include WEST~\cite{Yu2024}, BerkeleyGW~\cite{DelBen2020}, and Yambo~\cite{Sangalli2019} for GW/BSE calculations, GPU-accelerated coupled-cluster methods~\cite{Dutta2023,ByteQC2025}, and density-matrix renormalization group applications~\cite{Menczer2024}.
These diverse efforts demonstrate both the broad applicability of GPU acceleration to quantum chemistry and condensed matter physics, and the technical challenges of achieving performance portability across different computational methods and hardware architectures.
}

{
For electron-phonon Wannier interpolation in particular, GPU acceleration was pioneered by Cepellotti et al.~\cite{Cepellotti2022} in the Phoebe code, which demonstrated the feasibility of accelerating phonon and electron transport calculations on GPUs.
Subsequently, Liu et al.~\cite{Liu2023} achieved substantial single-node speedups by exploiting the nested-loop structure of the interpolation over uniform wavevector grids and carefully optimizing data reuse within this framework.
}

Here, we demonstrate successful porting of \epw{} to GPUs, and highlight the performance improvements on three different HPC systems: Vista at the Texas Advanced Computing Center (TACC), Aurora at the Argonne Leadership Computing Facility (ALCF), and Perlmutter at the National Energy Research Scientific Computing Center (NERSC). Of these systems, Vista and Perlmutter use NVIDIA accelerators, while Aurora uses Intel accelerators.

Our main goal is to accelerate the interpolation of \eph{} matrix elements, which constitutes the core module of \epw{} and underpins all the downstream modules of the code.
We follow a porting strategy aligned with that of the Quantum ESPRESSO (QE) suite~\cite{QE2009,QE2017,QEGPU}, which employs OpenACC~\cite{OpenACC} and OpenMP~\cite{OpenMP} offloading models, combined with vendor-optimized numerical libraries~\cite{Carnimeo2023,Ruffino2024}. 
We demonstrate how this scheme, together with OpenMP multithreading, can seamlessly be integrated into the existing message passing interface (MPI) parallelism, and provides a performance-portable implementation on both NVIDIA and Intel GPUs.

{
In contrast to prior GPU implementations of \eph{} interpolation, which primarily focused on either single-node performance using alternative loop orderings or vendor-agnostic portability, the present work targets both for deployment on leadership-class HPC systems. Our implementation combines GPU acceleration with multi-level MPI parallelization and OpenMP threading in a unified approach that enables strong scalability to thousands of nodes. By adopting directive-based OpenACC and OpenMP programming, the implementation achieves performance portability across GPU architectures while remaining fully integrated into the EPW workflow, with strict numerical equivalence to the CPU-only version.
Importantly, the framework is production-ready and immediately usable by the EPW user community, enabling large-scale electron-phonon calculations for end users, as demonstrated by our phonon-limited transport calculations on a topological material containing approximately one hundred atoms, a scale that has been considered infeasible so far.
}

\bigskip
\noindent\textbf{RESULTS}

\smallskip
\noindent\textbf{{Computational aspects of electron-phonon matrix interpolation}}

\smallskip

\noindent
The main quantity needed for a quantitative description of \eph{} interactions in materials is the \eph{} matrix,
denoted as $g_{mn\nu}(\kk,\qq)$~\cite{GiustinoRMP2017}:
\begin{equation}
    \label{eq:g_blo}
    g_{mn\nu}(\kk,\qq) = \int_{\Omega} u^*_{m\kk+\qq}(\rr) \, \Delta_{\qq\nu} \vscf(\rr) \, u_{n\kk}(\rr) \, d\rr,
\end{equation}
which represents the probability amplitude for a Bloch state with cell-periodic wavefunction $u_{n\kk}(\rr)$, momentum $\kk$, and band index $n$ to scatter to another state $u_{m\kk+\qq}(\rr)$ with momentum $\kk+\qq$ and band index $m$ via a variation in the self-consistent Kohn-Sham potential $\Delta_{\qq\nu} \vscf(\rr)$ arising from a phonon with momentum $\qq$ and mode index $\nu$.
Here, the perturbing potential is written as
\begin{align}
    \label{eq:dv_scf}
    &\Delta_{\qq\nu} \vscf(\rr) \nonumber \\
    &= \sum_{\kappa}^{\Na} \sum_{\alpha}^{3} \sum_{\RRp}^{\NRRp} \mathrm{u}_{\ka,\nu}(\qq) \, \frac{\pd \vscf}{\pd \tau_{\ka,\RRp}}(\rr) \, e^{-i\qq\cdot(\rr-\RRp)},
\end{align}
where $\kappa$ is the atomic site index that runs over the $\Na$ atoms in the unit cell, 
$\alpha$ labels the three cartesian coordinates ($x$, $y$, and $z$), 
$\RRp$ is the lattice vector that spans the $\NRRp$ unit cells in the Born-von K\'arm\'an (BvK) supercell,
$\tau_{\ka,\RRp}$ is the $\alpha$ component of the cartesian coordinates of the $\kappa$-th atom in the $\RRp$ unit cell, 
and $\mathrm{u}_{\ka,\nu}(\qq)$ is the phonon mode eigenvector.
We note that $u_{n\kk}(\rr)$ and $u_{m\kk+\qq}(\rr)$ are normalized in the primitive unit cell of volume $\Omega$.

The direct evaluation of all quantities in Eqs.~\eqref{eq:g_blo} and \eqref{eq:dv_scf} with DFT and DFPT is possible in principle but challenging in practice, especially when $g_{mn\nu}(\kk,\qq)$ needs to be computed on highly dense $\kk$ and $\qq$ grids in the Brillouin zone.
For example, in \textit{ab initio} transport calculations for bulk silicon, the $\kk$- and $\qq$-point grids required for converged mobility can be as dense as $100 \times 100 \times 100$~\mbox{\cite{Ponce2021}}.

This difficulty can be overcome by representing the \eph{} interaction in terms of Wannier functions~\cite{GiustinoPRB2007}:
\begin{equation}
    \label{eq:g_wan}
    g_{m'n'\kappa\alpha}(\RRe,\RRp) 
    =\int_V w^*_{m'}(\rr) \, \frac{\pd \vscf}{\pd \tau_{\ka,\RRp}}(\rr) \, w_{n'}(\rr-\RRe) \,d\rr,
\end{equation}
where $w_{m'}(\rr)$ represent a Wannier function localized in the home unit cell, and $w_{n'}(\rr-\RRe)$ is a one localized in the unit cell labeled by the lattice vector $\RRe$.
Here, the Wannier functions are normalized with respect to a BvK supercell of volume $V$.

The Wannier functions and Bloch states are related via a generalized Fourier transform (FT)~\cite{Wannier1937,Marzari1997,SouzaPRB2001,Marzari2012}:
\begin{align}
    \label{eq:wf_blo}
    {w_{n'}}(\rr-\RRe)
    &= \frac{1}{\NRRe} \sum_{n}^{\Nb} \sum_{\kkp}^{\Nkkp} {\psi_{n\kkp}}(\rr) \, U_{nn'\kkp} \, e^{-i\kkp\cdot\RRe}, \\
    \label{eq:bf_wan}
    {\psi_{n\kk}}(\rr) 
    &= \sum_{n'}^{\Nb} \sum_{\RRe}^{\NRRe} {w_{n'}}(\rr-\RRe) \, U^*_{nn'\kk} \, e^{i\kk\cdot\RRe}.
\end{align}
Here, $\NRRe$ is the number of unit cells in the BvK supercell, $\Nb$ is the number of bands included in the Wannierization procedure, $\Nkkp$ is the number of coarse-grid $\kkp$ points, ${\psi_{n\kkp}}(\rr)$ denotes the Bloch state with band index $n$ and wave vector $\kkp$, and the unitary matrix $U_{nn'\kkp}$ connects the Bloch and Wannier gauges.

Similarly, the Wannier representation of the \eph{} interaction, $g_{m'n'\kappa\alpha}(\RRe,\RRp)$, can be transformed to and from the Bloch representation $g_{mn\nu}(\kk,\qq)$:
\begin{align}
    \label{eq:g_coarse}
    &g_{m'n'\ka}(\RRe,\RRp) = \frac{1}{\NRRe \NRRp}
        \sum_{\kkp}^{\Nkkp} \sum_{\qqp}^{\Nqqp} \sum_{m}^{\Nb} \sum_{n}^{\Nb} \sum_{\nu}^{\Nm} g_{mn\nu}(\kkp,\qqp) \nonumber \\
        &\qquad\times U^*_{mm'\kkp+\qqp} \, U_{nn'\kkp} \, \textrm{u}^{-1}_{\nu,\ka}(\qqp) \, e^{-i(\qqp\cdot\RRp + \kkp\cdot\RRe)}, \\
    \label{eq:g_fine}
    &g_{mn\nu}(\kk,\qq) = \sum_{m'}^{\Nb} \sum_{n'}^{\Nb} \sum_{\kappa}^{\Na} \sum_{\alpha}^{3} \sum_{\RRe}^{\NRRe} \sum_{\RRp}^{\NRRp} g_{m'n'\ka}(\RRe,\RRp) \nonumber \\
    &\qquad\times U_{mm'\kk+\qq} \, U^*_{nn'\kk} \, \mathrm{u}_{\ka,\nu}(\qq) \, e^{i(\qq\cdot\RRp + \kk\cdot\RRe)},
\end{align}
where $\Nqqp$ is the number of coarse-grid $\qqp$, $\Nm$ is the number of phonon modes, which is equal to $3\Na$.
Here, $g_{mn\nu}(\kkp,\qqp)$ in Eq.~\eqref{eq:g_coarse} is obtained from DFT and DFPT calculations using Eqs.~\eqref{eq:g_blo} and \eqref{eq:dv_scf}.

Equation~\eqref{eq:g_fine} is referred to as the Wannier interpolation of the \eph{} matrix~\cite{GiustinoPRB2007}.
A key to understanding why this generalized FT can be useful lies in the spatial localization of the matrix elements in the Wannier representation.
When maximally localized Wannier functions~\cite{Marzari2012} are employed, $g_{m'n'\ka}(\RRe,\RRp)$ [Eq.~\eqref{eq:g_wan}] decays rapidly with respect to $|\RRp|$ and $|\RRe|$~\cite{GiustinoPRB2007}.
As a result, in the calculation of $g_{mn\nu}(\kk,\qq)$ via Eq.~\eqref{eq:g_fine}, the sum over $\RRp$ and $\RRe$ can be truncated with minimal loss of accuracy.

In the EPW workflow, the cutoff for the truncation over $\RRe$ and $\RRp$ is determined by the Wigner-Seitz (WS) cell of a BvK supercell; in turn, the size of the BvK supercell is set by the coarse $\kkp$ and $\qqp$ grids used to calculate $g_{m'n'\kappa'\alpha'}(\RRe,\RRp)$.
The first step of the workflow is to compute $g_{m'n'\kappa'\alpha'}(\RRe,\RRp)$ with relatively coarse grids of $\kkp$ and $\qqp$ using Eq.~\eqref{eq:g_coarse}, and the resulting $g_{m'n'\kappa'\alpha'}(\RRe,\RRp)$ is called the \eph{} matrix in the Wannier representation.
The second step is the \eph{} matrix interpolation,
whereby $g_{mn\nu}(\kk,\qq)$ is calculated using Eq.~\eqref{eq:g_fine} over fine grids of $\kk$ and $\qq$.
For this reason, $g_{mn\nu}(\kk,\qq)$ obtained from the interpolation step is usually referred to as the fine-grid \eph{} matrix.
In Eq.~\mbox{\eqref{eq:g_fine}}, we deliberately denote the interpolated fine-grid \mbox{\eph{}} matrix as \mbox{$g_{mn\nu}(\kk,\qq)$} to distinguish it from the \mbox{$g_{mn\nu}(\kkp,\qqp)$} appearing in Eq.~\mbox{\eqref{eq:g_coarse}}; this choice is to emphasize that the former can be evaluated for arbitrary dense, fine \mbox{$\kk$} and \mbox{$\qq$} grids.

Accurate interpolation of the fine-grid \eph{} matrix requires calculating $g_{m'n'\ka}(\RRe,\RRp)$ for sufficiently large electron and phonon WS supercells.
Typically, the numbers of $\RRe$ and $\RRp$ vectors within such WS cells range from 10$^2$ to 10$^3$.
In the case of bulk silicon, for example, coarse $\kkp$ and $\qqp$ grids of size $8\times8\times8$ and $4\times4\times4$, respectively, can provide an accurate fine-grid interpolation for transport coefficients.
In this example, the numbers of the electron and phonon WS vectors are $\NRRe = 617$ and $\NRRp = 93$, respectively, and the entire $g_{m'n'\ka}(\RRe,\RRp)$ can be stored in the main memory of a computing node as a double-precision complex array of size 0.4~GB (assuming $\Nb = 8$ and $\Nm = 6$).
The size of this buffer is much smaller than that of the high bandwidth memory (HBM) on typical GPU nodes in HPC clusters.
This \textit{key observation} forms the basis of our GPU porting strategy for the \eph{} matrix interpolation.

\bigskip

\noindent\textbf{Analysis of computational complexity}

\smallskip

\noindent
To develop an efficient GPU porting strategy, we analyze the computational complexity of two different algorithms that implement the evaluation of Eq.~\eqref{eq:g_fine}. Nearly all computationally intensive features in \epw{} involve the evaluation of $g_{mn\nu}(\kk,\qq)$ on uniform fine $\kk$ and $\qq$ grids, therefore we focus on this scenario.

\begin{figure}
    \centering
    \includegraphics{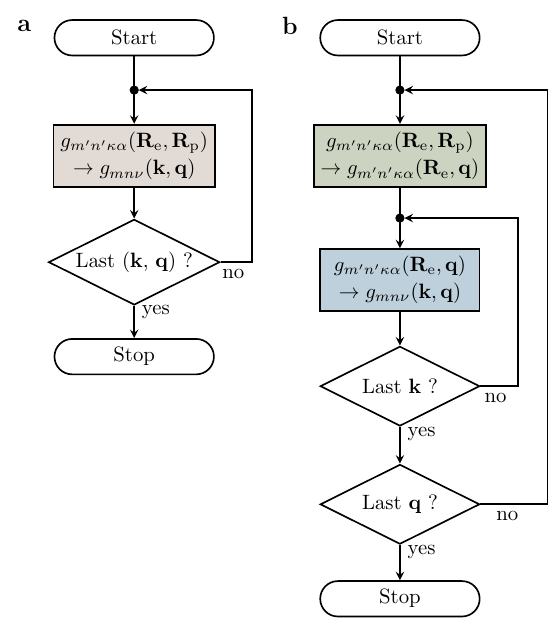}
    \caption{
        {\bf Simplified flowcharts for electron-phonon matrix interpolation.} {\bf a} Single-loop scheme, where the interpolation of the fine-grid matrix $g_{mn\nu}(\kk,\qq)$ from the coarse-grid matrix $g_{m'n'\ka}(\RRe,\RRp)$ is performed by repeating a single-step procedure [Eq.~\eqref{eq:g_fine}] over $\kk$ and $\qq$ pairs; {\bf b} Nested-loop scheme, where the interpolation is divided into two steps [Eqs.~\eqref{eq:g_fine_1} and \eqref{eq:g_fine_2}].  These substeps are carried out at different levels in the nested loop structure: outer and inner loops for $\qq$ and $\kk$, respectively.  This approach requires to allocate a buffer array for storing the intermediate result $g_{m'n'\ka}(\RRe,\qq)$.  The size of this buffer is smaller than that of the coarse-grid matrix by a factor of $\Nqq$, the number of $\qq$ points.
    }
    \label{fig:intp_loop}
\end{figure}

The simplest approach to evaluate Eq.~\eqref{eq:g_fine} is the single-loop algorithm shown in Fig.~\ref{fig:intp_loop}{\bf a}.
For fine $\kk$ and $\qq$ grids of size $N_\kk$ and $N_\qq$, respectively, the single-loop implementation evaluates Eq.~\eqref{eq:g_fine} for all $\Nkk \times \Nqq$ pairs of $\kk$ and $\qq$.
The number of double-precision complex multiplications required by this na\"ive algorithm is
\begin{align}
    \label{eq:Tsing}
    \Teph^{\rm SL}
    &= (\underbrace{\Nb^2 \Nm \NRRe \NRRp}_{\substack{\textrm{FT w.r.t. }\RRp}} + \underbrace{\Nb^2 \Nm \NRRe}_{\textrm{FT w.r.t. }\RRe} \nonumber \\
    &\quad\quad+ \underbrace{2\Nb^2\Nm}_{\substack{\textrm{Rotation by}\\
    U_{mm'\kk+\qq},U^*_{nn'\kk}}} +  \underbrace{\Nb^2 \Nm^2}_{\substack{\textrm{Rotation by}\\ \mathrm{u}_{\ka,\nu}(\qq)}} ) \, \Nkk \Nqq.
\end{align}
This can be considered to be an estimate for the computational workload required by the single-loop algorithm.
Hereafter, we measure the arithmetic cost in terms of the number of double-precision complex multiplications, which dominate the floating-point operation counts.
We further note that this estimate assumes that a complex array of size $\Nb^2\Nm$
was pre-allocated and used as a buffer during the interpolation.
The buffer stores intermediate results from the matrix-vector and matrix-matrix multiplications in Eq.~\eqref{eq:g_fine}, more specifically, those from a successive multiplication of $e^{i\qq\cdot\RRp}$, $e^{i\kk\cdot\RRe}$, $U_{mm'\kk+\qq}$, $U^*_{nn'\kk}$, and $\mathrm{u}_{\ka,\nu}(\qq)$ with $g_{m'n'\ka}(\RRe,\RRp)$.

The workload required by the entire \eph{} interpolation step can be significantly reduced by using the nested-loop algorithm shown in Fig.~\ref{fig:intp_loop}{\bf b}.
This hierarchical scheme exploits the fact that, in the case of uniform grids, the same set of $\kk$ needs to be looped over for each $\qq$.

In the nested-loop algorithm, the evaluation of Eq.~\eqref{eq:g_fine} is further divided into two parts:
\begin{align}
    \label{eq:g_fine_1}
    &g_{m'n'\ka}(\RRe,\qq)
        = \sum_{\RRp}^{\NRRp} g_{m'n'\ka}(\RRe,\RRp) \, e^{i\qq\cdot\RRp}, \\
    \label{eq:g_fine_2}
    &g_{mn\nu}(\kk,\qq)
        = \sum_{m'}^{\Nb} \sum_{n'}^{\Nb} \sum_{\kappa}^{\Na} \sum_{\alpha}^3 \sum_{\RRe}^{\NRRe} g_{m'n'\ka}(\RRe,\qq) \nonumber \\
        &\quad\quad\quad\quad\quad\quad\times U_{mm'\kk+\qq}\, U^*_{nn'\kk} \, \mathrm{u}_{\ka,\nu}(\qq) \, e^{i\kk\cdot\RRe}.
\end{align}
The first part [Eq.~\eqref{eq:g_fine_1}] performs the FT of the coarse-grid matrix $g_{m'n'\ka}(\RRe,\RRp)$ with respect to the phonon WS vectors $\RRp$.
This is carried out only once at the beginning of each $\qq$ iteration (the outer loop in Fig.~\ref{fig:intp_loop}{\bf b}),
and the result $g_{m'n'\ka}(\RRe,\qq)$ is stored in a buffer for the subsequent steps within the $\qq$ loop.
The size of the temporary array is negligible compared to that of the coarse-grid \eph{} matrix.
The second part [Eq.~\eqref{eq:g_fine_2}] applies, successively, the FT with respect to the electron WS vectors, $\RRe$,
and the unitary rotations with respect to electron bands $m'$ and $n'$ and phonon indices $\kappa$ and $\alpha$,
to the intermediate quantity $g_{m'n'\ka}(\RRe,\qq)$ from the previous step.
For each $\qq$, Eq.~\eqref{eq:g_fine_2} is evaluated over the fine $\kk$ grid (the inner loop in Fig.~\ref{fig:intp_loop}{\bf b}).
The resulting computational workload of the nested-loop algorithm is:
\begin{align}
    \label{eq:Tnest}
    &\Teph^{\rm NL}= \underbrace{\Nb^2 \Nm \NRRe \NRRp}_{\textrm{FT w.r.t. }\RRp} \Nqq + ( \underbrace{\Nb^2 \Nm \NRRe}_{\textrm{FT w.r.t. }\RRe} + \underbrace{2\Nb^2\Nm}_{\substack{\textrm{Rotation by} \\ U_{mm'\kk+\qq},U^*_{nn'\kk}}} \nonumber \\ 
    &\qquad\qquad\qquad\qquad\qquad~~+\underbrace{\Nb^2 \Nm^2}_{\substack{\textrm{Rotation by} \\ \mathrm{u}_{\ka,\nu}(\qq)}}) \, \Nkk \Nqq.
\end{align}
For bulk systems, $\Nb$, $\Nm$, $\NRRe$ and $\NRRp$ typically fall in the range of 1--40, 3--30, $10^2$--$10^3$ and $10^2$--$10^3$, respectively.
Taking bulk silicon as an example, where $\Nb = 8$, $\Nm = 6$, $\NRRe = 617$, and $\NRRp = 93$, 
the estimated ratio between the workloads of the single-loop and the nested-loop algorithms is approximately:
\begin{align}
    \label{eq:Tratio}
    \frac{\Teph^{\rm NL}}{\Teph^{\rm SL}}
    &\approx
        \frac{1}{\Nkk}
        + \frac{1}{\NRRp} \left(1 + \frac{2}{\NRRe} + \frac{\Nm}{\NRRe}\right),
\end{align}
where $\Teph^{\rm SL} \approx \Nb^2\Nm\NRRe\NRRp\Nkk\Nqq$ was used, since the other terms in \mbox{Eq.~\eqref{eq:Tsing}} are at least two orders of magnitude smaller.
Therefore, in large-scale calculations where $\Nkk \gg \NRRp$, the ratio $\Teph^{\rm NL}/\Teph^{\rm SL}$ is as small as $\sim$1/93.
This analysis shows that the nested-loop algorithm is much more efficient than the single-loop case. This is the strategy traditionally used in \epw{}, until v5.9~\cite{LeeNPJ2023}.

\bigskip

\noindent\textbf{Refactoring of parallelism}

\bigskip

\noindent\textbf{Two-level MPI parallelism}. 
Traditionally, \epw{} has used a single-level MPI scheme to distribute the workload of the \eph{} matrix interpolation.
Until \epw{} 5.9, the single MPI communicator, called the pool parallelization, was solely responsible for accelerating the computationally intensive steps in the nested-loop algorithm.

The flat MPI scheme is inherently efficient in parallelizing the $\kk$ loop (the inner loop in Fig.~\ref{fig:intp_loop}{\bf b}) over the evaluation of Eq.~\eqref{eq:g_fine_2}, especially in large-scale calculations, since the number of iterations scales with the fine $\kk$-grid size $\Nkk$.

By contrast, this parallelization strategy is less effective for the evaluation of Eq.~\eqref{eq:g_fine_1}.
Typically, the number of phonon WS vectors $\NRRp$ is smaller than the fine $\kk$-grid size $\Nkk$, which limits the efficiency of pool parallelization over $\RRp$.
Additionally, the actual implementation of the evaluation of Eq.~\eqref{eq:g_fine_1} involves MPI collective communications and input/output (IO) operations.
This technique substantially reduces the runtime memory footprint but introduces additional {MPI+IO} overheads at each $\qq$ iteration.
For these reasons, \epw{} 5.9 had an early saturation issue in the speed-up with respect to the number of pools $\Npo$ in large-scale calculations ($\Nkk,\Nqq \gg 1$)~\cite{LeeNPJ2023}.

A simple yet realistic estimate of the computational workload per MPI rank that takes the above considerations into account can be written as:
\begin{align}
    \label{eq:WL_EPW_5.9}
    \Teph^\text{\epw{} 5.9} &= \Bigg\{ N_{\rm pool}^{\textrm{MPI+IO}} + \bigg[ \underbrace{ \frac{\Nb^2\Nm\NRRe\NRRp}{\Npo}}_{\substack{\RRp~\textrm{distribution} \\ \textrm{across pools}}} \nonumber \\ &\qquad+ \underbrace{ \frac{\Nb^2\Nm ( \NRRe+2+\Nm )\Nkk}{\Npo}}_{\substack{\kk~\textrm{distribution}\\\textrm{across pools}}}
        \bigg] \Bigg\} \Nqq,
\end{align}
where $N_{\rm pool}^{\textrm{MPI+IO}}$ represents the MPI+IO overhead cost per pool, converted in units of double-precision complex multiplications, i.e., $N_{\rm pool}^{\textrm{MPI+IO}} = T_{\rm pool}^{\textrm{MPI+IO}}/T^{\text{zmul}}$ with $T_{\rm pool}^{\textrm{MPI+IO}}$ being the wall time for the MPI+IO per pool and $T^{\text{zmul}}$ the time required for a single double-precision complex multiplication.
Equation~\eqref{eq:WL_EPW_5.9} indicates that, for given $N_{\rm CPU}$ cores, the workload per rank is minimized when $\Npo=N_{\rm CPU}$, since this enables maximal distribution of the evaluation of Eqs.~\eqref{eq:g_fine_1} and \eqref{eq:g_fine_2} across pools.
However, the parallelization efficiency of this one-to-one mapping strategy is restricted by the MPI+IO overhead.

In large-scale calculations where a number of CPU cores are required, {$N_{\rm pool}^{\textrm{MPI+IO}}$ is expected to dominate the other contributions in the bracket in Eq.~\eqref{eq:WL_EPW_5.9}.
We denote the ratios of these other terms to the contribution $N_{\rm pool}^{\textrm{MPI+IO}}$ as
\begin{align}
    x_1 = \frac{\Nb^2\Nm\NRRe\NRRp}{N^{\text{MPI+IO}}_{\rm pool}},~
    x_2 = \frac{\Nb^2\Nm(\NRRe+2+\Nm)\Nkk}{N^{\text{MPI+IO}}_{\rm pool}}.
\end{align}
In the scenario where all available CPU cores are fully utilized within the single-level MPI scheme of EPW 5.9, i.e., $\Npo = N_{\rm CPU}$, the estimated workload is
\begin{align}
    \Teph^{\text{EPW 5.9}} = N^{\text{MPI+IO}}_{\rm pool} \left(1 + \frac{x}{N_{\rm CPU}}\right)\Nqq,
\end{align}
where $x=x_1 + x_2$.
For sufficiently large $N_{\rm CPU}$, $\Teph^{\textrm{\epw{} 5.9}}$ is dominated by the MPI+IO overhead, and saturates to $N_{\rm pool}^{\textrm{MPI+IO}}\Nqq$.
This factor cannot be reduced by increasing $N_{\rm CPU}$; empirically, it is found to be relatively insensitive to $N_{\rm CPU}$ at small values of $N_{\rm CPU}$, and to increase slowly at large $N_{\rm CPU}$.
This effect explains the early saturation of the speedup in strong-scaling tests that is commonly observed in the single-level MPI scheme of \epw{} 5.9~\cite{LeeNPJ2023}.}

\begin{figure}
    \centering
    \includegraphics{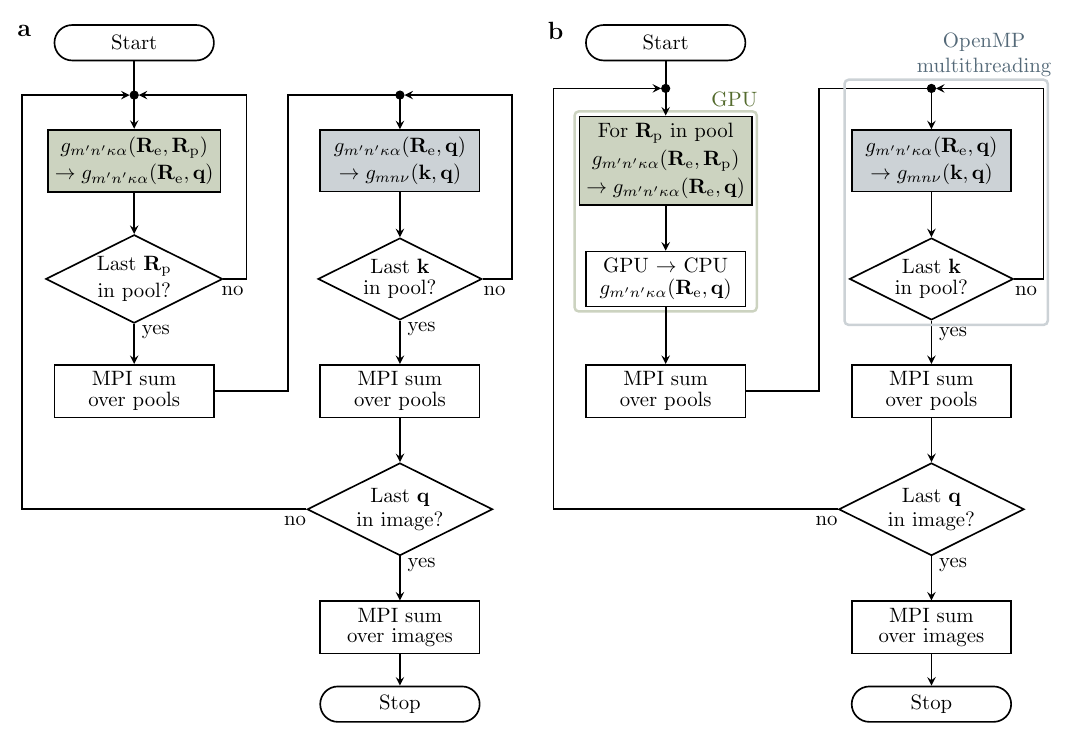}
    \caption{
        {\bf Flowcharts for electron-phonon matrix interpolation implemented in EPW.}
        {\bf a} Two-level MPI scheme for the interpolation of the electron–phonon matrix $g_{mn\nu}(\kk,\qq)$, implemented in \epw{} 6.0.
        This approach builds on the nested-loop algorithm (Fig.~\ref{fig:intp_loop}{\bf b}), where the interpolation is performed in two substeps (green and blue boxes).
        The workload is distributed across the image and pool parallelization levels: the $\RRp$ and $\kk$ indices are divided among pools, and collective MPI reductions (e.g., summation) over pools are required to assemble the $\qq$ slice of the intermediate $g_{m'n'\ka}(\RRe,\qq)$ and the final $g_{mn\nu}(\kk,\qq)$ arrays.
        The $\qq$ points are distributed across images, and a final MPI summation over images gives the complete $g_{mn\nu}(\kk,\qq)$ for all $\qq$ points.
        {\bf b} Hybrid MPI--GPU--OpenMP scheme, introduced in \epw{} 6.1.
        GPU acceleration and OpenMP multithreading are incorporated into the outer $\qq$ loop of the two-level MPI framework.
        In this design, the Fourier transform that converts $g_{m'n'\ka}(\RRe,\RRp)$ to $g_{m'n'\ka}(\RRe,\qq)$ (green box) is offloaded to GPUs for speed, while OpenMP threads allow each MPI rank to fully exploit the available CPU cores.
    }
    \label{fig:mpi_nested_loop}
\end{figure}

To address this issue, \epw{} recently added support for an additional MPI communicator, which is referred to as the image parallelization.
Figure~\ref{fig:mpi_nested_loop}{\bf a} illustrates how the resulting two-level MPI scheme naturally integrates with the nested-loop algorithm.
The role of the pool parallelization remains the same as before, while the image parallelization enables distribution of the $\qq$ loop (the outer loop in Fig.~\ref{fig:mpi_nested_loop}{\bf a}) over the evaluation of Eqs.~\eqref{eq:g_fine_1} and \eqref{eq:g_fine_2} across images. 
Our estimate for the workload per MPI rank in this case is
\begin{align}
    \label{eq:WL_EPW_6.0}
    \Teph^\text{\epw{} 6.0} &= \overbrace{\Bigg\{ N_{\rm pool}^{\textrm{MPI+IO}} + \bigg[ \underbrace{ \frac{\Nb^2\Nm\NRRe\NRRp}{\Npo} }_{\substack{\RRp~\textrm{distribution} \\ \textrm{across pools}}} \qquad\qquad\qquad}^{\qq~\textrm{distribution across images}} \nonumber \\ &\qquad+ \underbrace{ \frac{\Nb^2\Nm ( \NRRe+2+\Nm )\Nkk}{\Npo} }_{\substack{\kk~\textrm{distribution} \\ \textrm{across pools}}}
        \bigg] \Bigg\} \frac{\Nqq}{\Nim}.
\end{align}
The appearance of $\Nqq/\Nim$ in Eq.~\eqref{eq:WL_EPW_6.0} reflects the fact that image parallelization introduces an additional level of parallelism.

In this two-level MPI scheme, maximal utilization of CPU resources is achieved when $N_{\rm CPU}=\Npo\Nim$.
In this case, the estimated workload is
\begin{align}
    {\Teph^{\text{EPW 6.0}}
    = N^{\text{MPI+IO}}_{\rm pool} \left(\frac{1}{\Nim} + \frac{x}{N_{\rm CPU}}\right) \Nqq.}
\end{align}
There is freedom to choose $\Nim$, with the constraint that it should divide $N_{\rm CPU}$.
This additional flexibility in workload distribution can result in significant performance improvements, particularly in large-scale calculations where the scaling efficiency of the single-level MPI scheme was largely limited by the MPI+IO overhead.
In this large $N_{\rm CPU}$ limit, where {$x/N_{\rm CPU} \ll 1$}, the ratio of the workload per MPI rank between \epw{} 5.9 and 6.0 can be estimated as
{
\begin{align}
    \label{eq:WL_ratio_1}
    \frac{ \Teph^{\textrm{\epw{} 6.0}} }{ \Teph^{\textrm{\epw{} 5.9}} }
    \approx
    \frac{1}{\Nim} 
    + \left(1-\frac{1}{\Nim}\right) \frac{x}{N_{\rm CPU}}.
\end{align}
It is noteworthy that, although $\Teph^{\text{EPW 6.0}}/\Teph^{\text{EPW 5.9}}$ monotonically decreases as $\Nim$ increases for a fixed $N_{\rm CPU}$, using the maximum possible number of images, i.e., $\Nim=N_{\rm CPU}$, does not necessarily lead to the best overall performance.
This is because the total wall time also includes pre- and post-processing routines, and they make the scaling deviate from the ideal $1/\Nim$ behavior predicted by Eq.~\eqref{eq:WL_EPW_6.0}.
Usually, for a given $N_{\rm CPU}$, there is an optimal $\Nim$ that minimizes the total wall time.
With $\Nim$ fixed to this optimal value, the second term in Eq.~\eqref{eq:WL_ratio_1} becomes negligible compared to the first term, as $x\Nim/N_{\rm CPU} \ll 1$.}
Consequently, Eq.~\eqref{eq:WL_ratio_1} reduces to $\Teph^{\textrm{\epw{} 6.0}} / \Teph^{\textrm{\epw{} 5.9}} \approx 1/\Nim$, which corresponds to an $\Nim$-fold speedup.
This speedup is mainly due to the reduction of the MPI+IO overhead cost from the image parallelization.

\bigskip

\noindent\textbf{GPU acceleration}.
The central idea of the present GPU acceleration strategy is that a single GPU outperforms a dozen CPU cores in computing the first part of the \eph{} matrix interpolation in the nested-loop algorithm [Eq.~\eqref{eq:g_fine_1}].
Instead of evaluating the sum via an explicit loop over $\RRp$, as it is done in \epw{} 5.9 and 6.0, this step can be carried out more efficiently by using the general matrix-vector product (GEMV) routine from the Basic Linear Algebra Subprograms (BLAS), with the matrix and vector being the coarse-grid matrix $g_{m'n'\ka}(\RRe,\RRp)$ and the Fourier phase factors $e^{i\qq\cdot\RRp}$~\cite{Liu2023}, respectively.
Acceleration is then achieved by offloading this GEMV operation using vendor-optimized linear algebra libraries such as cuBLAS (NVIDIA), oneMKL (Intel), and rocBLAS (AMD).

It is well known that GEMV has low arithmetic intensity, which makes it memory-bound rather than compute-bound.
In general, offloading such memory-bound kernels to GPUs is not considered best practice.
This is because, in contrast to the general matrix–matrix product (GEMM), which is compute-bound, GEMV cannot fully exploit the peak floating-point performance of GPUs.
Still, GEMV can largely benefit from the high memory bandwidth of GPUs, and therefore offloading GEMV to GPUs can provide significant acceleration over CPUs.

In fact, the memory bandwidth of modern GPUs exceeds that of CPUs, as shown by Tab.~\ref{tab:gpu_node}, where we summarize the hardware specifications of the GPU-accelerated and CPU-only computing nodes used for the present tests.
For example, a single CPU node in Perlmutter (NERSC) has two AMD EPYC 7763 processors, whereas a GPU node is equipped with one AMD EPYC 7763 processor and four NVIDIA A100 GPUs.
The peak memory bandwidth of AMD EPYC 7763 is 0.2~TB/s per CPU, while NVIDIA A100 provides 1.6~TB/s per GPU.
Consequently, the total memory bandwidth of a GPU node (6.4~TB/s) is sixteen times greater than that of a CPU node (0.4~TB/s).

To fully exploit the high memory bandwidth of GPUs, frequent data transfers between CPU and GPU should be avoided.
This condition is naturally satisfied in the Wannier interpolation of the \eph{} matrix elements,
because the same coarse-grid \eph{} matrix $g_{m'n'\ka}(\RRe,\RRp)$ can be reused in the evaluation of Eqs.~\eqref{eq:g_fine_1} and \eqref{eq:g_fine_2} throughout the calculation.
As a result, once data in the Wannier
representation have been uploaded from the CPU, they can remain in the GPU's main memory for the entire interpolation step.

One challenge with this offloading strategy is to handle cases where the coarse-grid \eph{} matrix is too large to fit into the GPU's main memory. We overcome this difficulty by leveraging the two-level MPI scheme, and we distribute the coarse-grid representation across pools:
a full copy of $g_{m'n'\ka}(\RRe,\RRp)$ is duplicated across images, and each copy is partitioned at the pool parallelization level with respect to $\RRp$.
In this scheme, each pool accesses only a portion of the coarse-grid \eph{} matrix throughout the calculation.
Before entering the interpolation, each pool uploads the partial copy of $g_{m'n'\ka}(\RRe,\RRp)$ to the GPU it is bound to.
The pool then carries out the first step of the \eph{} matrix interpolation [Eq.~\eqref{eq:g_fine_1}] with respect to its partition of $\RRp$.
Following this step, the partial result is transferred back to the CPU,
and the full $g_{m'n'\ka}(\RRe,\qq)$ is constructed through a collective communication across pools.

An important byproduct of this strategy is the elimination of the MPI+IO step associated with the evaluation of Eq.~\eqref{eq:g_fine_1}.
This simplification significantly improves the performance of the \eph{} matrix interpolation in large-scale calculations by removing the associated MPI+IO overhead.
We note that this scheme can readily be implemented by using the OpenACC and OpenMP interoperability with vendor-optimized math libraries; the implementation details are provided in the Methods section.

\bigskip

\noindent\textbf{OpenMP multithreading}.
A common practice for utilizing multiple GPUs in distributed environments is to bind each MPI rank to a single GPU.
This scheme is simple yet often effective,
particularly when the implementation does not require frequent data transfers between CPU and GPU.
Since typical GPU nodes in supercomputers have 1--6 GPUs per node,
this one-to-one binding strategy sets an upper limit on the number of MPI ranks per node.
In turn, this bound limits the maximum number of pools that can be used at runtime. 
As a result, fewer CPU cores can be used for parallelizing the $\kk$ loop over pools.

To overcome this challenge, we employ a hybrid MPI-GPU-OpenMP strategy, as shown in  Fig.~\ref{fig:mpi_nested_loop}{\bf b}.
In this scheme, the inner loop is distributed across pools as in the MPI-only approach,
but each MPI rank can spawn multiple OpenMP threads to parallelize its $\kk$ iterations.

Taking into account both GPU acceleration and OpenMP multithreading,
the computational workload per MPI rank can be estimated as
\begin{align}
    \label{eq:wl_rank_3}
    \Teph^\text{\epw{ 6.1}} &= \overbrace{
        \bigg[\underbrace{\frac{\Nb^2 \Nm \NRRe\NRRp}{\alpha_{\rm GPU} \Npo}}_{\substack{\RRp~\textrm{distribution across pools}\\+~\textrm{GPU acceleration}}}\qquad\qquad\qquad}^{\qq~\textrm{distribution across images}} \nonumber \\
        &\qquad+ \underbrace{ \frac{\Nb^2\Nm(\NRRe+2+\Nm)\Nkk}{\Npo \Nth}}_{\substack{\kk~\textrm{distribution across pools}\\+~\textrm{OpenMP multithreading}}}
        \bigg] \frac{\Nqq}{\Nim},
\end{align}
where $\Nth$ is the number of OpenMP threads per MPI rank, $\alpha_{\rm GPU}$
represents the acceleration provided by the GPU for the evaluation of Eq.~\eqref{eq:g_fine_1} (a larger value means faster execution),
while the denominator $\Npo \Nth$ reflects that the inner loop is parallelized across pools and OpenMP threads.

{Given the total number of CPU cores $N_{\rm CPU}$ in a fixed number of nodes, with $\Nth = N_{\rm CPU}/(\Nim \Npo)$ being the number of OpenMP threads per MPI rank, the workload $\Teph^{\text{EPW 6.1}}$ can be written as
\begin{align}
    \Teph^{\text{EPW 6.1}} = N^{\text{MPI+IO}}_{\rm pool} \left(\frac{y\,x_1 + x_2}{N_{\rm CPU}} \right) \Nqq,
\end{align}
where $y$ denotes $\Nth/\alpha_{\rm GPU}$.
For typical thread counts $\Nth$ of 4--18 and an empirically determined $\alpha_{\rm GPU}$ of 27--33, $y$ is smaller than or comparable to 1.
Subject to these constraints, the ratio of the estimated wall time between \epw{} 5.9 and 6.1 is
\begin{align}
    \label{eq:WL_ratio_2}
    \frac{ \Teph^{\textrm{\epw{} 6.1}} }{ \Teph^{\textrm{\epw{} 5.9}} }
    \approx \frac{y\,x_1 + x_2}{N_{\rm CPU}}.
\end{align}
From Eq.~\eqref{eq:WL_ratio_1}, we also see that 
\begin{align}
    \label{eq:WL_ratio_3}
    \frac{ \Teph^{\textrm{\epw{} 6.1}} }{ \Teph^{\textrm{\epw{} 6.0}} }
    \approx \frac{\frac{(yx_1 + x_2)\Nim}{N_{\rm CPU}}}{1 + \left(1 - \frac{1}{\Nim} \right)\frac{x\Nim}{N_{\rm CPU}}}.
\end{align}}
{This factor remains small whenever the strong-scaling saturation occurs: the numerator is negligible, i.e., $(yx_1+x_2)\Nim/N_{\rm CPU} \ll 1$, while the denominator is close to unity, since $x\Nim/N_{\rm CPU} \ll 1$ [see also Eq.~\eqref{eq:WL_ratio_1} and the discussion below]. This guarantees an additional speed-up, arising from the GPU acceleration ($y=\Nth/\alpha_{\rm GPU}$) and the reduced MPI communication costs incorporated into the hybrid MPI-GPU-OpenMP scheme.}

\bigskip

\begin{figure}
    \centering
    \includegraphics{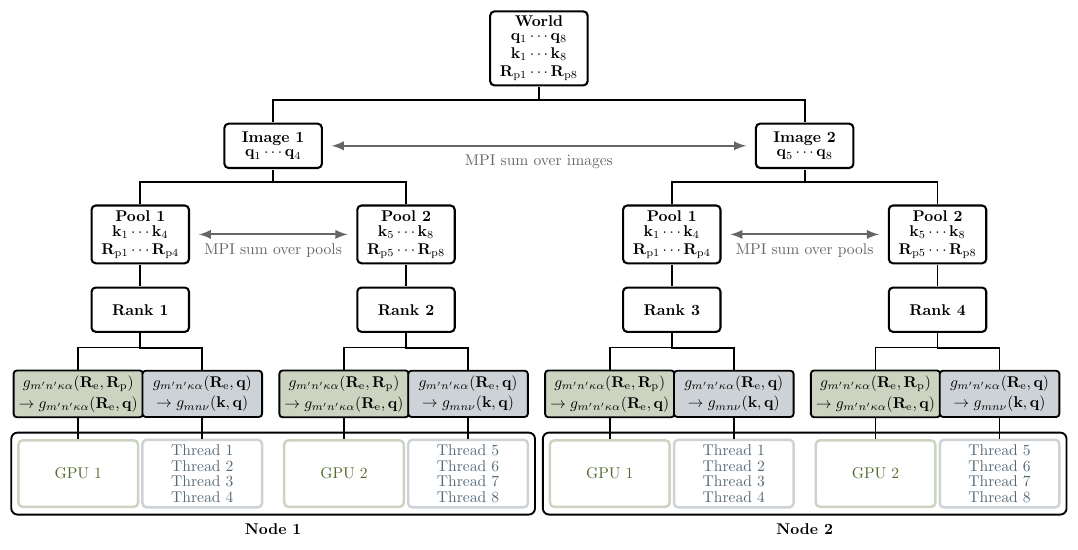}
    \caption{
        \textbf{Overview of the workload distribution in the hybrid MPI-GPU-OpenMP scheme.}
        Each computing node in this example system has two GPUs and eight hardware threads (CPU cores).
        The workload is distributed with one image per node, two pools per image, one GPU per pool, and four OpenMP threads per pool.
    }
    \label{fig:work_dist}
\end{figure}

\noindent\textbf{Workload distribution}.
Figure~\ref{fig:work_dist} shows an overview of the workload distribution within the two-level MPI-GPU-OpenMP scheme in \epw{} 6.1.
The figure illustrates how the \eph{} matrix interpolation is distributed for fine $\kk$ and $\qq$ grids of size $\Nkk = 8$ and $\Nqq = 8$, respectively, and for a coarse-grid \eph{} matrix of size $\NRRp = 8$.
Here, we consider a hypothetical system consisting of two computing nodes, each of which is equipped with two GPUs and eight hardware threads (physical CPU cores).
Following the one-to-one binding scheme, four MPI ranks are employed in total.
The MPI ranks are divided into two images, and each image is further split into two pools.

The first step of the \eph{} matrix interpolation [Eq.~\eqref{eq:g_fine_1}] is offloaded to GPUs.
After the GPU result is transferred back to the CPUs,
a collective summation across pools is carried out to complete the step.
The next step is the $\kk$ loop over the evaluation of Eq.~\eqref{eq:g_fine_2}, which is distributed across pools.
In this example, each pool is responsible for four $\kk$ points.
The MPI rank associated with each pool then spawns four OpenMP threads for parallelization over CPU cores.
Once the $\kk$ loop finishes, an inter-pool collective summation is performed to finalize the step.

At the top level of this hierarchy, the total workload is distributed across images.
We note that no inter-image communication is required during the outer loop ($\qq$).
Only a single inter-image collective summation is required after the outer loop, to collect the partial results from the images.
This design minimizes MPI communication overhead, and enables near-ideal scalability, as we demonstrate in the following section.

{
\smallskip
\noindent\textbf{Practical guidance}.
To obtain optimal performance, users should first assess the size of the coarse-grid \eph{} matrix, $g_{m'n'\ka}(\RRe,\RRp)$, relative to the available GPU memory.
When the entire coarse-grid matrix fits within the GPU memory, one image per GPU without pool parallelization is recommended, with OpenMP threading employed for parallelizing the inner $\mathbf{k}$ loop.
When the coarse-grid matrix exceeds the memory capacity of a single GPU, pool parallelization should be enabled to distribute the object across multiple GPUs.

Because \eph{} interpolation is embarrassingly parallel over $\mathbf{q}$ points, the total wall time scales nearly linearly with the number of GPU nodes and can be estimated as $N_q/N_{\rm image}\times T_{1q}$, 
where $N_q$ is the number of fine $\mathbf{q}$-points, $N_{\rm image}$ is the number of images employed, and $T_{1q}$ is the wall time for a single-$\mathbf{q}$ calculation with the number of $\mathbf{k}$ points fixed at the full calculation, which can be obtained from a short test calculation.

Strong scaling is primarily limited by the cost of loading the coarse-grid \eph{} matrix into CPU and GPU memory; for moderate matrix sizes (less than approximately 10~GB), file I/O speed (a few GB/s) dominates, making large node counts inefficient for calculations with wall times of only a few minutes.
Weak scaling over $\mathbf{q}$-points is inherently ideal due to the embarrassingly parallel nature of the $\mathbf{q}$-loop, while increasing $\mathbf{k}$-grid density is addressed through intra-node parallelization via pools, OpenMP threading, and GPU acceleration.
For target grid selection, we recommend converging physical observables with respect to grid density: typical semiconductors reqiure grids of $80 \times 80 \times 80$ to $400 \times 400 \times 400$~\cite{LeeNPJ2023}, while metals or systems with sharp Fermi surface features may require a denser grid; the energy window parameter significantly reduces the effective number of contributing $\mathbf{k}$-points for semiconductors and insulators.
The current workload distribution assigns $\mathbf{q}$-point Fourier transforms to GPUs and $\mathbf{k}$-point transforms to OpenMP threads. This design is efficient in typical bulk materials where symmetry operations and energy windowing reduce the effective number of $\mathbf{k}$-points. For systems with flat bands near the Fermi level, where many $\mathbf{k}$-points contribute within a given energy window, reducing the energy window parameter simultaneously decreases both the number of contributing $\mathbf{k}$-points and bands. As demonstrated below for zigzag stanene nanoribbons, a challenging test case with $N_k \approx N_q$, the framework still achieves approximately two-fold acceleration over the two-level MPI-only scheme, demonstrating robustness even in such demanding scenarios.
}

{
\smallskip
\noindent\textbf{Numerical precision}.
The new implementation for \eph{} matrix interpolation employs double-precision arithmetic throughout the code base, and no mixed-precision optimizations were used.
We have verified that the CPU and GPU code paths produce numerically identical results when provided with the same coarse-grid \eph{} matrix data, thereby ensuring full numerical reproducibility.
}

\bigskip

\noindent\textbf{Benchmarks}

\smallskip

\noindent\textbf{Single-node performance}.
We carried out single-node benchmarks on the Perlmutter, Vista, and Aurora supercomputers.
For each machine, we tested three different versions of \epw{} (5.9, 6.0, and 6.1) on the available combinations of CPU architectures and GPU accelerators.
In each case, we selected the numbers of pools and images (when applicable) 
that yield the best performance, including pre- and post-processing procedures as well as \eph{} matrix interpolation.
{For detailed information about the benchmark systems and optimal parallelization parameters, see the Methods section.}

\begin{figure}
    \includegraphics{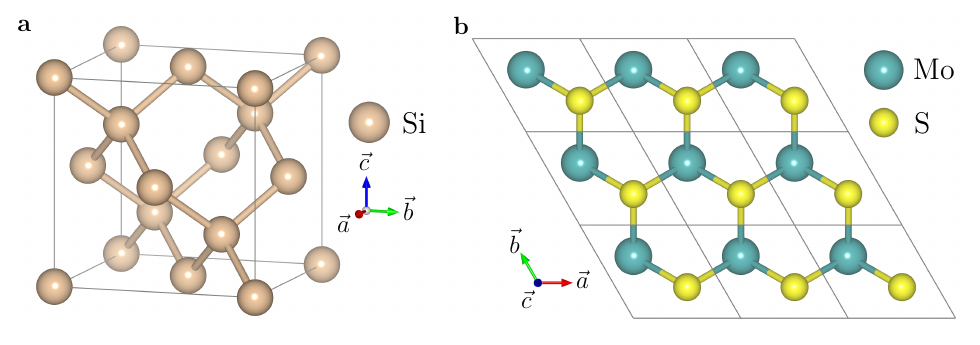}
    \caption{
        \textbf{Crystal structure of systems used in benchmark calculations.} {\bf a} Bulk silicon (conventional cell); {\bf b} Two-dimensional MoS$_2$ ($3\times3\times1$ supercell). 
        The figures were generated with the VESTA program~\cite{Momma2008}.
    }
    \label{fig:structure}
\end{figure}

The benchmark tests were performed for the electron mobility in bulk silicon (see Fig.~\ref{fig:structure}{\bf a} for the crystal structure) at room temperature (300~K), using the {\it ab initio} Boltzmann transport module in \epw{}~\cite{Ponce2018,Ponce2021,LeeNPJ2023}.
This test is representative of a scenario where the \eph{} matrix interpolation dominates the total computation time.
The coarse $\kk'$ and $\qq'$ grids were $8\times8\times8$ and $4\times4\times4$, respectively, which correspond to $\NRRe = 617$ electron WS vectors and $\NRRp = 93$ phonon WS vectors.
The \eph{} matrix interpolation was performed on much denser grids, with both fine $\kk$ and $\qq$ grids of size $96\times96\times96$.
We note that the number of actual $\kk$ and $\qq$ used in the calculations differs from $96^3$,
since two additional optimization steps are performed in \epw{} to reduce computational cost.
In the first step, a pre-screening of $\kk$ and $\qq$ points is used, skipping the calculation of the \eph{} matrix elements involving states outside a 0.3~eV energy window near the conduction band minimum, which do not contribute to the mobility.
In the second step, only $\kk$ points within the irreducible Brillouin zone are considered~\cite{LeeNPJ2023}.
After these optimizations, the effective $\Nkk$ and $\Nqq$ were approximately 1,000 and 50,000, respectively.
These effective values of $\Nkk$ and $\Nqq$ are what actually determine the workload distributed across images, pools, and GPUs in the two-level MPI scheme.

\begin{figure}
    \centering
    \includegraphics{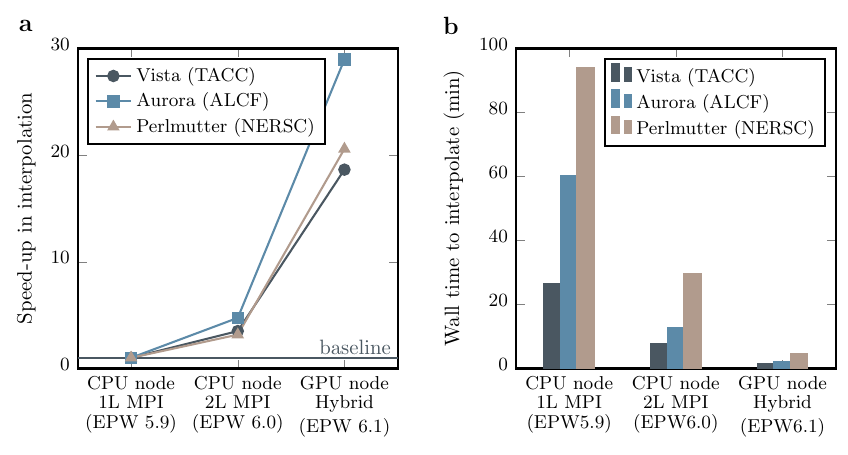}
    \caption{
        \textbf{Comparison of single-node performance of electron-phonon matrix interpolation.}
        {\bf a} Relative speedup and {\bf b} wall time in minutes, measured on three supercomputers: Vista at the Texas Advanced Computing Center (TACC), Aurora at the Argonne Leadership Computing Facility (ALCF), and Perlmutter at the National Energy Research Scientific Computing Center (NERSC).
        Benchmarks are based on {\it ab initio} Boltzmann transport calculations for bulk silicon.
        For each system, the baseline calculation was performed on a single CPU node using the single-level MPI scheme (1L MPI, \epw{} 5.9). The performance improvements of the two-level MPI (2L MPI, \epw{} 6.0) and the hybrid MPI-GPU-OpenMP (Hybrid, \epw{} 6.1) schemes are reported relative to this baseline.
        { }
    }
    \label{fig:bm_single}
\end{figure}

{
Figure~\ref{fig:bm_single} summarizes the single-node benchmark results, illustrating the progression of performance improvements from \epw{} 5.9 through 6.0 to 6.1.
The first stage of improvement, from \epw{} 5.9 to 6.0, introduces two-level MPI parallelization over both $\qq$ images and $\kk$ pools.
This yields performance gains of 3.1--4.7$\times$, which can be understood in terms of the MPI+IO overhead model in Eq.~\eqref{eq:WL_ratio_1}.
Importantly, this two-level parallelization is not merely an incremental improvement but serves as the essential foundation for scalable GPU acceleration: it enables the distribution of workloads across thousands of GPU nodes while fully utilizing available CPU resources through the hybrid MPI-OpenMP-GPU framework.
The second stage, from \epw{} 6.0 to 6.1, adds GPU offloading and OpenMP multithreading, providing additional speedups of 5.3--6.3$\times$ across Perlmutter, Vista, and Aurora (see Fig.~\ref{fig:bm_single}{\bf a}).
This level of GPU acceleration is on par with that achieved in GPU porting of the Quantum ESPRESSO suite~\cite{Carnimeo2023,Ruffino2024}.
Here, we highlight that the speedup achieved in the \eph{} matrix interpolation directly affects the wall time of almost all features in \epw{} that require dense sampling of \eph{} matrix elements.
Combined, the full progression from \epw{} 5.9 to 6.1 achieves overall speedups of 19--29$\times$, highlighting both the effectiveness of the algorithmic restructuring and the performance portability of the hybrid scheme.}
{As shown in Fig.~\ref{fig:bm_single}{\bf b}, the absolute wall time drops below 5 minutes with GPU acceleration and OpenMP multithreading.}
{While there exists some variability in the measured wall time, newer systems tend to exhibit smaller wall times, below 2 minutes.}

It is worth pointing out that these improvements in \epw{} 6.1 hold across a variety of hardware and software environments, including CPU architectures (x86 and Arm), GPU vendors (NVIDIA and Intel), compilers (NVIDIA HPC SDK and Intel OneAPI), and linear algebra libraries (cuBLAS and OneMKL).
We also anticipate comparable improvements on AMD GPUs, because the performance of rocBLAS is expected to be similar to other vendor-optimized BLAS libraries, and our hybrid MPI-GPU-OpenMP scheme leverages their peak performance by design.

\smallskip

\noindent\textbf{Multi-node scalability}.
Beyond performance portability on single GPU nodes, exascale calculations demand scalability on large numbers of nodes. To assess the scalability of the present implementation, we consider two use cases.

The first scenario is representative of a mid-scale calculation.
We used the Boltzmann transport calculation for bulk silicon (Fig.~\ref{fig:structure}{\bf a}) from the single-node performance benchmark, but with a slight increase in the size of fine $\kk$ and $\qq$ grids ($120\times120\times120$) as compared to the single-node benchmark.
The benchmark tests were performed on Vista, using up to 32 GPU nodes.
The workload was distributed with 8 images per node, one pool per image, and 9 OpenMP threads per pool.

The second scenario is representative of a large-scale calculation on thousands of GPU nodes.
We performed a calculation of the phonon-induced electron self-energy of two-dimensional MoS$_2$ (Fig.~\ref{fig:structure}{\bf b}), starting from coarse $\kkp$ and $\qqp$ grids with $12\times12\times1$ and $6\times6\times1$ points, respectively, and interpolating to ultra dense fine $\kk$ and $\qq$ grids with $380\times380\times1$ points. The benchmark tests were performed on Aurora, using up to 1,024 GPU nodes, corresponding to 6,144 GPUs. The workload was distributed with {1 image per node, 12 pools per image}, and 8 OpenMP threads per pool.

\begin{figure}
    \centering
    \includegraphics{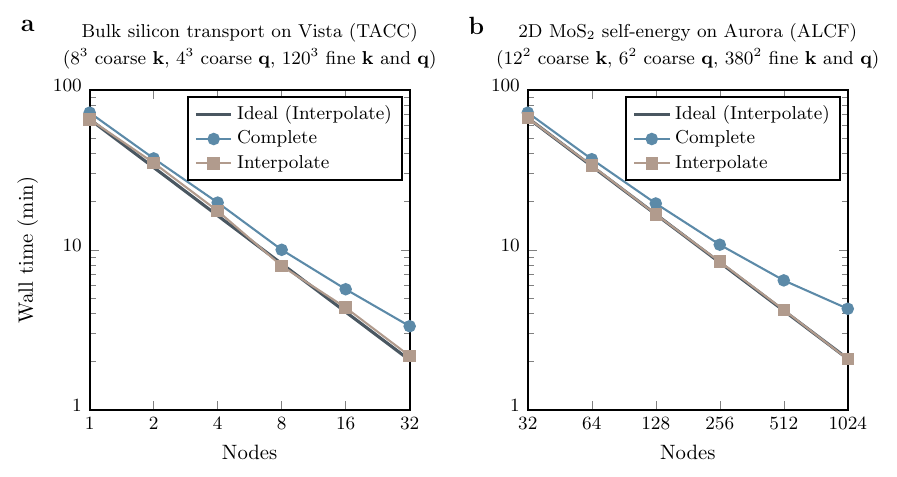}
    \caption{
        \textbf{Multi-node scalability of the hybrid two-level MPI-GPU-OpenMP strategy (\epw{} 6.1).}
        {\bf a} Mid-scale benchmark of an {\it ab initio} Boltzmann transport calculation for bulk silicon, performed on up to 32 GPU nodes on Vista [Texas Advanced Computing Center (TACC)]. Each node employed 8 MPI ranks with 9 OpenMP threads per task, distributed across 8 images and one pool per image.
        {\bf b} Large-scale benchmark of an electron self-energy calculation for monolayer MoS$_2$, performed on 32--1024 GPU nodes on Aurora [Argonne Leadership Computing Facility (ALCF)]. Each node used 12 MPI ranks, each bound to one GPU tile and 8 CPU threads, with { 1 image per node}, { 12 pools per image}, and 8 OpenMP threads per pool.
    }
    \label{fig:bm_multi}
\end{figure}

For both scenarios, we measured both the total wall time and the wall time for Wannier interpolation of the \eph{} matrix. Figure~\ref{fig:bm_multi} shows near-perfect linear scaling in the \eph{} matrix interpolation for both mid-scale and large-scale calculations on Vista and Aurora (brown squares in Fig.~\ref{fig:bm_multi}). Such a good scaling reflects the embarrassingly parallel nature of the nested-loop algorithm for the Wannier interpolation of the \eph{} matrix elements, see Fig.~\ref{fig:intp_loop}{\bf b}.
{The two-level MPI parallelization eliminates inter-node communication during \eph{} interpolation: once the coarse-grid data are distributed, each node proceeds independently. The observed near-ideal scaling from 32 to 1,024 GPU nodes confirms that communication overheads are negligible. In passing, by Amdahl's law, scaling from smaller allocations (2--32 nodes) is expected to be equally good.
We performed scaling tests using 1--4 nodes and confirmed this point.
}

The blue circles in Fig.~\ref{fig:bm_multi} show the wall time required to complete these calculations.
These data include pre-processing and post-processing in addition to the \eph{} interpolation. Also in this case we observe very good scaling, with speedups of 21.6$\times$ and 16.8$\times$ for the mid-scale test on Vista and the large-scale test on Aurora, respectively. Importantly, at the largest node counts these advanced calculations required less than 5 minutes.

These results demonstrate that \epw{} 6.1, which is being released simultaneously with the present study, is highly efficient and is ready for exascale calculations of \eph{} couplings.

\smallskip

\begin{figure}
    \centering
    \includegraphics{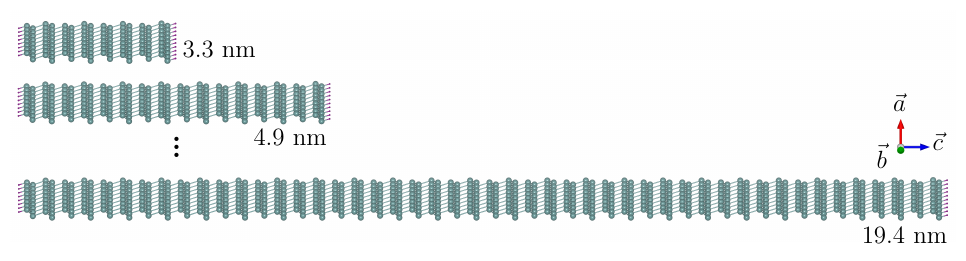}
    \caption{
        \textbf{Crystal structures of hydrogen-passivated zigzag stanene nanoribbons (ZSNRs) with varying widths.}
        Three representative widths are shown: 3.3~nm (top), 4.9~nm (middle), and 19.4~nm (bottom).
        The calculations span widths from 3.3 to 19.4~nm, corresponding to 18 to 98 atoms per unit cell.
        The hydrogen-passivated zigzag edges are along the crystallographic $\vec{b}$ axis; $1 \times 8 \times 1$ supercells are shown here.
        In our first-principles calculations, a vacuum of 12~\AA{} was added along the $\vec{a}$ and $\vec{c}$ axes.
        The figure was generated with the VESTA program~\cite{Momma2008}.
    }
    \label{fig:zsnr_widths}
\end{figure}

\begin{figure}
    \centering
    \includegraphics{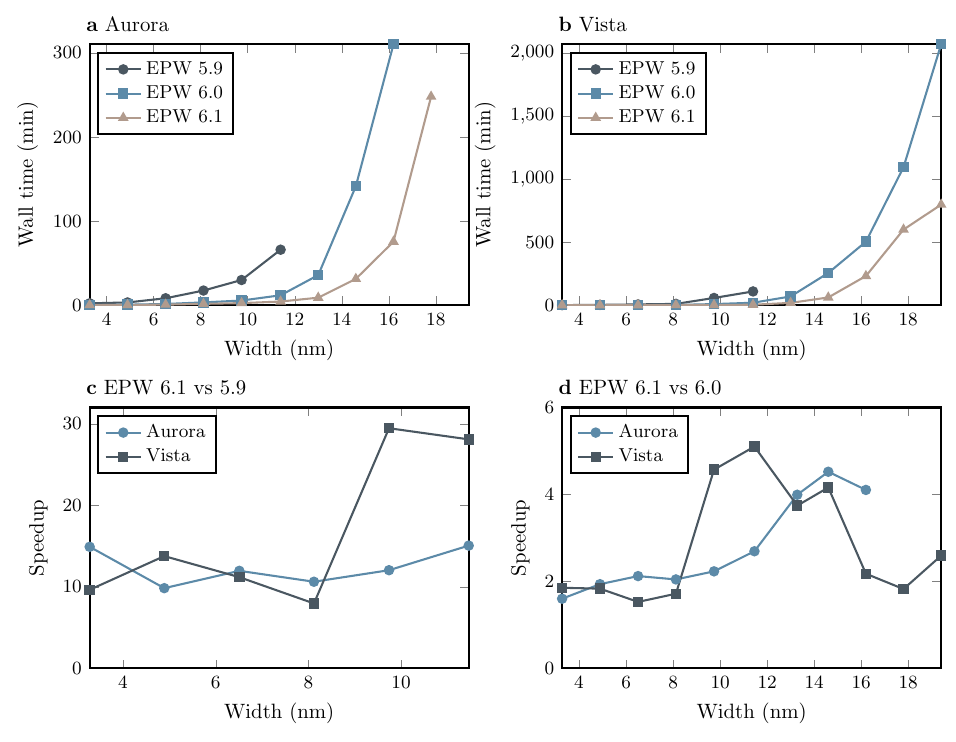}
    \caption{
        \textbf{Wall time and relative speedup comparison between \epw{ 5.9}, \epw{ 6.0}, and \epw{ 6.1}
        for hydrogen-passivated zigzag stanene nanoribbon calculations.}
        {\bf a},\,{\bf b} Wall time as a function of nanoribbon width comparison between different versions of EPW on Aurora [Argonne Leadership Computing Facility (ALCF)] and Vista [Texas Advanced Computing Center (TACC)].
        Wall times shown are for the end-to-end \eph{} interpolation workflow (excluding the coarse-grid calculation step).
        For smaller ribbon widths, single-node calculations were performed.
        For larger widths where multiple nodes were required, the wall time shown is the measured value multiplied by the number of nodes used, enabling direct comparison with the single-node results.
        {\bf c},\,{\bf d} Relative speedups of EPW 6.1 with respect to EPW 5.9 and EPW 6.0, respectively.
    }
    \label{fig:zsnr_time}
\end{figure}

{
\noindent\textbf{Large-scale application: phonon-limited edge channel transport}.
Finally, we present a first-principles study of phonon-limited carrier transport in large-scale systems containing up to a hundred atoms per unit cell, focusing on the influence of finite size effects on the underlying \eph{} physics enabled by our MPI-GPU-OpenMP framework.
As a representative example, we consider hydrogen-passivated zigzag stanene nanoribbons (ZSNRs), a candidate material for nanoelectronics that hosts edge conduction states arising from the topological nature of monolayer tin films~\cite{XuPRL2013,Fu2017,Deylgat2022}.
Although transport in ZSNRs has been investigated using nonequilibrium Green’s function methods~\cite{LiPRB2024}, a systematic \textit{ab-initio} Boltzmann transport study on phonon-limited carrier transport as a function of the ribbon width has been lacking.
This has largely been due to the prohibitive computational cost of evaluating \eph{} matrix elements for such large unit cells.
Overcoming this limitation, we provide a physically transparent reference for phonon-limited transport in ZSNRs, which can be naturally extended to study more complex phenomena arising from magnetic ordering and different edge passivation~\cite{Fu2017,Deylgat2022,LiPRB2024}.

Figure~\ref{fig:zsnr_widths} shows the atomic structure of hydrogen-passivated ZSNRs for three representative ribbon widths: 3.3~nm, 4.9~nm, and 19.4~nm.
The zigzag edges are passivated by hydrogen atoms along the $\vec{b}$ axis, which suppresses dangling-bond states at the edges.
Here, the ZSNR width is defined as the separation with respect to the $\vec{c}$ axis between the hydrogen atoms terminating the two opposite zigzag edges.
We note that the widest ribbon, with a width of 19.4~nm, contains 98 atoms in the unit cell.

Table~\ref{tab:gsize_zsnr} summarizes the calculation parameters that determine the primary device-memory constraint, which is the size of the coarse-grid \eph{} matrix for ZSNRs of different widths.
As the ribbon width increases, the size of the coarse-grid \eph{} matrix grows rapidly, from 1.9~GB for the narrowest ribbon (3.3~nm, 18 atoms) to 458~GB for the widest case (19.4~nm, 98 atoms).
This exponential growth in size is not only due to the increasing number of electronic and phonon bands with increasing system size, but also to the increase in the number of coarse-grid $\kk$ points, which was essential to achieve an accurate Wannier interpolation of the electronic bands in this system.
In particular, an insufficient number of coarse-grid $\kk$ points led to poor Wannier interpolation, resulting in unphysical band crossings across the bulk gap.
Since accurate Wannierization near the Fermi level is a prerequisite for reliable \eph{} interpolation and subsequent transport calculations, this requirement significantly increases the memory footprint of the workflow.
As a result, the coarse-grid \eph{} matrix for the largest system (458~GB) far exceeds not only the high-bandwidth memory available on a single GPU, but also the main memory capacity of a computing node on most HPC systems, highlighting the importance of developing a scalable implementation for \eph{} matrix interpolation.

Figure~\ref{fig:zsnr_time} illustrates the rapid increase in wall time required to complete the \textit{ab-initio} Boltzmann transport workflow as the ZSNR width increases. The figure reports the single-node wall time for the \eph{} matrix interpolation step as a function of ribbon width, comparing \epw{} versions 5.9, 6.0, and 6.1 on the Aurora and Vista systems. ZSNRs constitute a particularly demanding test case, as their low symmetry requires treating nearly all $\kk$ and $\qq$ points explicitly, placing the calculation in a regime where the numbers of $\kk$ and $\qq$ points are comparable.
Despite this unfavorable workload balance, the GPU-accelerated \epw{} 6.1 consistently achieves the shortest wall times across all ribbon widths and on both platforms, enabling calculations that would otherwise be prohibitively expensive or impractical.
The resulting speedup relative to \epw{} 5.9 ranges from approximately 7$\times$ to 30$\times$, depending on the system size and computing platform.

It is also noteworthy that the scaling with respect to ribbon width differs between the two machines. In particular, a sudden relative performance degradation of \epw{}~6.0 compared to \epw{}~6.1 is observed when the ribbon width exceeds approximately 8~nm on Vista and 14~nm on Aurora. This behavior originates from the memory constraints inherent to the \epw{}~6.0 implementation, which are dominated by the size of the buffer array used to store intermediate \eph{} matrix elements, $g_{m'n'\ka}(\RRe,\qq)$.
In \epw{}~6.0, where only (two-level) MPI parallelization is available, this buffer must be allocated independently by every MPI rank. Consequently, for systems with a large number of electronic bands and phonon modes, the number of MPI ranks that can be placed on a node can be severely limited.

The impact of this limitation depends strongly on the hardware architecture. Aurora provides approximately four times more main memory per node than Vista.
Moreover, on Vista the CPU nodes have substantially larger memory capacity than the GPU nodes, and each GPU node hosts only a single GPU.
As a result, once the ribbon width reaches approximately 14~nm, the coarse-grid \eph{} matrix no longer fits within the memory of a single GPU node on Vista. Although distributing the coarse-grid matrix over the pool-parallelization level was designed to accommodate this regime, an inevitable loss of efficiency arises due to the unfavorable CPU/GPU memory ratio on this system.
By contrast, on Aurora, where the GPU memory per node is significantly larger and comparable to the CPU memory, reflecting the design trend of recent exascale systems, the relative performance gap between \epw{}~6.1 and \epw{}~6.0 increases more gradually with ribbon width.

Figure~\ref{fig:band_comparison} presents the electronic band structures of hydrogen-passivated ZSNRs for ribbon widths ranging from 3.3~nm to 16.2~nm. 
In all cases, a characteristic linear band crossing appears near the Brillouin-zone boundary, originating from topological edge states inherited from the quantum spin Hall insulator phase of monolayer stanene. 
Because of the one-dimensional nature of the system, the corresponding density of states (DOS) exhibits pronounced singularities at the band extrema. 
In particular, the high DOS associated with the edge states leads to an effective pinning of the Fermi level, which is located approximately 25~meV above the band-crossing (Dirac) point for all ribbon widths considered. 
In addition, the quality of the Wannierization is evident from the excellent agreement between the DFT and Wannier-interpolated bands, which essentially overlap throughout the energy range of interest.
This level of agreement justifies the size of the coarse-grid \eph{} matrix employed in the calculations, as summarized in Table~\ref{tab:gsize_zsnr}, and confirms the reliability of the subsequent \eph{} interpolation and transport analysis.

To further elucidate the relationship between the electronic structure and transport properties, Figure~\ref{fig:zsnr_dos} presents the phonon-limited electrical conductivity at three representative temperatures, 50, 100, and 300~K, together with the electronic density of states (DOS) shown as a background, for the 3.3~nm hydrogen-passivated ZSNR.
A clear correlation between the DOS and the conductivity is observed, which is most pronounced at the lowest temperature (50~K).
In particular, singular features in the DOS are reflected as local minima (dips) in the conductivity as a function of the Fermi level.

As the temperature increases from 50~K to 100~K, the conductivity decreases, consistent with the typical metallic behavior, where the reduction of the relaxation time with increasing temperature dominates the temperature dependence of the conductivity.
Interestingly, a qualitatively different behavior emerges between 100~K and 300~K, where the conductivity at the Fermi level increases with temperature. This behavior can be understood by considering the more general expression $\sigma = -e^2/(2\pi)^3 \sum_{n} \int d\kk \, (\pd f^0_{n\kk}/\pd\varepsilon_{n\kk}) \, v_{n\kk}^2 \,\tau_{n\kk}$, where the thermal broadening of the derivative of the Fermi-Dirac distribution allows additional contributions from the electronic states far away from the Fermi level.
At 100~K ($\sim$8.7~meV), the broadening becomes comparable to the energy separation between the Fermi level and the local maximum of the edge-state bands, corresponding to the DOS peak slightly above the Fermi level, as can be seen in Fig.~\ref{fig:zsnr_dos}.
As a result, states with high DOS near the band crossing increasingly contribute to the conductivity at higher temperatures. 
We note that this nontrivial temperature dependence is confined to a narrow energy window around the Fermi level, which is itself pinned by the high DOS arising from the topological edge states, and is therefore a distinctive feature of the present system.

Figure~\ref{fig:sigma_combined}{\bf a} shows that, for all ribbon widths considered, the temperature dependence of the conductivity follows a similar qualitative trend: a crossover from conventional metallic behavior to a nontrivial regime arising from thermal broadening of the electronic occupations toward states with higher group velocities. For undoped, pristine hydrogen-passivated ZSNRs, we find that up to widths of approximately 20~nm the temperature dependence of the conductivity is governed by the pinning of the Fermi level caused by the high density of states associated with the edge bands.

As can be seen in Fig.~\ref{fig:sigma_combined}{\bf b}, 100~K marks a crossover temperature for ribbon widths larger than 8~nm, above which the width dependence of the conductivity changes.
As the ZSNR width increases, bulk-like bands move closer in energy to the edge states and begin to contribute more effectively to charge transport, leading to an overall increase in conductivity. 
By contrast, for smaller widths (3.3~nm and 4.9~nm), the conductivity decreases with increasing width, with a crossover occurring at approximately 6.5~nm. 
Below this width, bulk-like bands that could enhance the conductivity remain energetically far from the Fermi level on the scale of thermal broadening, as can be seen from Fig.~\ref{fig:band_comparison}, and therefore do not contribute significantly to transport.

We emphasize that obtaining a clear physical understanding of phonon-limited transport in systems of this size, containing on the order of 100 atoms per unit cell, has only become tractable as a result of our recent advances in $\qq$ parallelization and GPU acceleration of the coarse-grid electron-phonon matrix calculations. 
The present study demonstrates, through an explicit large-scale application, that such calculations can now be carried out in practice and yield physically transparent results. 
In particular, the strong sensitivity of the transport properties to microscopic details of the electronic band dispersion highlights the richness of the underlying physics that can be accessed with this approach. 
This opens the door to exploring more intricate phenomena, such as the interplay between magnetic ordering and spin-dependent band topology, including possible topological protection effects associated with spin degrees of freedom, which may provide valuable insights for future nanoscale spintronic device applications. 

In passing, we note that there are ongoing development efforts to extend the hybrid MPI-GPU-OpenMP framework to the construction of the coarse-grid \eph{} matrix, with the goal of enabling fully end-to-end acceleration of \eph{} workflows within the EPW code.
GPU acceleration of the coarse-grid stage proved essential for carrying out the widest nanoribbon case with a width of 19.4 nm, for which the size of the coarse-grid \eph{} matrix would otherwise have rendered the calculation impractical. These developments further highlight the importance of extending GPU acceleration beyond the interpolation stage to support large-scale, production-level electron-phonon studies.
}

\bigskip

\noindent\textbf{DISCUSSION}

\smallskip

In this work, we have presented a highly efficient and performance-portable GPU implementation of electron-phonon calculations using Wannier interpolation.
The new implementation achieves speed-ups of 19--29 times across several leadership-class supercomputers, including Perlmutter, Vista, and Aurora, equipped with NVIDIA and Intel GPUs,
{as compared to the single MPI parallelization scheme of EPW v5.9}.
Moreover, the present GPU implementation exhibits nearly ideal strong scalability up to thousands of GPU nodes on Aurora, one of the world's first exascale systems.
These advances enable {\it ab initio} studies of electron-phonon physics in
{complex systems, as we demonstrated here for large-width topological
stanene nanoribbons}.
We expect that this development will open new possibilities for high-throughput screening of materials for next-generation electronics, optoelectronics, and quantum technologies with \epw{}, and for generating large datasets for AI/ML workflows. More generally, we hope that the present 
{work will serve as a reference and benchmark for developing and testing related implementations}.

\bigskip
\bigskip
\bigskip

\noindent\textbf{METHODS}

\smallskip

\noindent\textbf{Implementation strategy}

\smallskip

\begin{figure}
    \centering
    \includegraphics{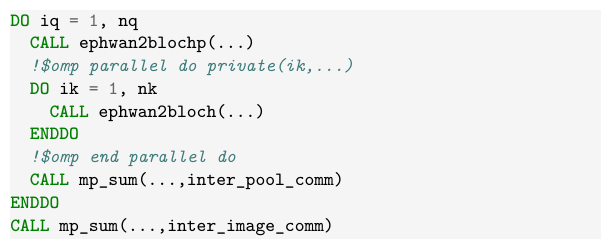}
    \caption{
        \textbf{Parallel code skeleton of the hybrid MPI-GPU-OpenMP scheme for the electron–phonon matrix interpolation implemented in \epw{} 6.1.}
        The workload for evaluating the fine-grid electron–phonon matrix $g_{mn\nu}(\kk,\qq)$ is distributed across two MPI communicators: images [outer $\qq$ loop, reduced with \texttt{mp\_sum(..., inter\_image\_comm)}] and pools [inner $\kk$ loop, reduced with \texttt{mp\_sum(..., inter\_pool\_comm)}].
        GPU acceleration is used in \texttt{ephwan2blochp}, which performs the Fourier transform over phonon Wigner–Seitz vectors $\RRp$ [Eq.~\eqref{eq:g_fine_1}], while OpenMP multithreading accelerates the evaluation of \texttt{ephwan2bloch} within the $\kk$ loop.
        \texttt{ephwan2bloch} performs the subsequent Fourier transform over electron Wigner–Seitz vectors $\RRe$ and evaluates Eq.~\eqref{eq:g_fine_2} for all $\kk$ points in each pool.
        See also Fig.~\ref{fig:mpi_nested_loop}{\bf b} for the corresponding schematic diagram.
    }
    \label{lst:main_loop}
\end{figure}

\noindent\emph{Workload distribution}.
Core subroutines of the \eph{} matrix interpolation in \epw{} were refactored to incorporate the hybrid MPI-GPU-OpenMP framework presented in this work.
Figure~\ref{lst:main_loop} presents a Fortran code skeleton that implements the key elements of this new parallel framework and reflects the flowchart in Fig.~\ref{fig:mpi_nested_loop}{\bf b}.
Here, the outer $\qq$ loop is distributed among images and reduced with \texttt{mp\_sum(..., inter\_image\_comm)}, while the inner $\kk$ loop is divided among pools and reduced with \texttt{mp\_sum(..., inter\_pool\_comm)}.
GPU acceleration is applied to \texttt{ephwan2blochp}, which carries out the Fourier transform over the phonon Wigner-Seitz vectors $\RRp$ [Eq.~\eqref{eq:g_fine_1}].
{The GPU-accelerated \texttt{ephwan2blochp} is free of collective IO operations such as \texttt{MPI\_FILE\_READ\_AT\_ALL}. The MPI communications across pools via \texttt{mp\_sum} are relatively inexpensive, since fewer pools, or even none, are required than in the CPU-only implementations (\epw{}~5.9 and 6.0).}
Within each pool, OpenMP multithreading is used to parallelize \texttt{ephwan2bloch}, which performs the subsequent Fourier transform over the electron Wigner--Seitz vectors $\RRe$ and evaluates Eq.~\eqref{eq:g_fine_2} for all $\kk$ points.
This OpenMP multithreading can easily be implemented with a single \texttt{parallel do} construct as shown in Fig.~\ref{lst:main_loop}.

\smallskip

\begin{figure}
    \centering
    \includegraphics{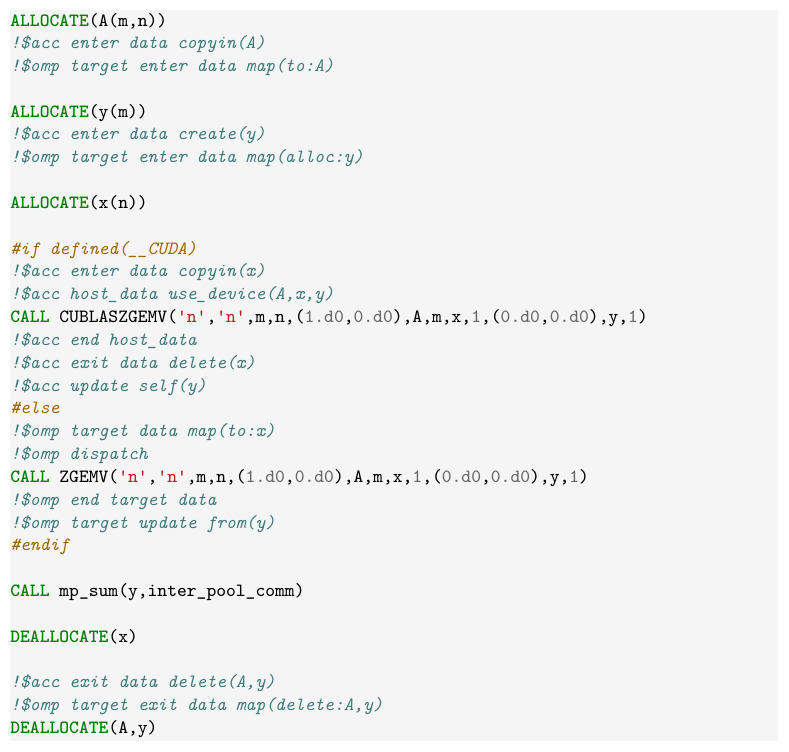}
    \caption{
        \textbf{Implementation snippet for offloading a complex general matrix-vector product (GEMV) to NVIDIA and Intel GPUs.}
        Offloading to NVIDIA and Intel accelerators is based on the OpenACC and OpenMP interoperability features from cuBLAS and oneMKL, respectively.
        The variables \texttt{A}, \texttt{x}, and \texttt{y} correspond to the coarse-grid electron–phonon matrix $g_{m'n'\ka}(\RRe,\RRp)$, the Fourier phase factor $e^{i\qq\cdot\RRp}$, and the intermediate result $g_{m'n'\ka}(\RRe,\qq)$, respectively [see Eq.~\eqref{eq:g_fine_1}].
    }
    \label{lst:zgemv_gpu}
\end{figure}

\noindent\emph{Portable GPU offloading of GEMV with OpenACC and OpenMP interoperability}. An important implementation aspect is how an expensive GEMV operation in \texttt{ephwan2blochp} is offloaded to GPUs.
The code pattern in Fig.~\ref{lst:zgemv_gpu} shows how this is accomplished as well as how the data creation and movement before and after the GEMV call to the GPU and the collective summation over pools are managed.
Here, \texttt{A}, \texttt{x}, and \texttt{y} correspond to $g_{m'n'\ka}(\RRe,\RRp)$, $e^{i\qq\cdot\RRp}$, and $g_{m'n'\ka}(\RRe,\qq)$, respectively.
The values of \texttt{A}, which represents the matrix $g_{m'n'\ka}(\RRe,\RRp)$, are initialized on the CPU side at the very beginning, and are distributed over $\RRp$ across pools.
Subsequently, \texttt{enter data copyin(A)} (OpenACC) or \texttt{target enter data map(to:A)} (OpenMP) are used for each pool to upload the partial $g_{m'n'\ka}(\RRe,\RRp)$ to the GPU.
On the other hand, the global array \texttt{y}, which represents the vector $g_{m'n'\ka}(\RRe,\qq)$, is left uninitialized at its allocation.
In this case, it is sufficient to use \texttt{enter data create(y)} (OpenACC) and \texttt{target enter data map(alloc:y)} (OpenMP), which create uninitialized  memory regions on the GPU.
For the vector \texttt{x}, which represents $e^{i\qq\cdot\RRp}$ and is declared as a local array and temporarily allocated within a subroutine call, we employ \texttt{data copyin(x)} (OpenACC) and \texttt{target data map(to:x)} (OpenMP) patterns, which encompass the GEMV call.

On NVIDIA GPUs, a GPU-accelerated GEMV can be performed using OpenACC's \texttt{host\_data} and \texttt{use\_device} directives.
The \texttt{use\_device(A,x,y)} clause exposes the device pointers of \texttt{A,x,y} to the \texttt{CUBLASZGEMV} routine in cuBLAS, which expects GPU memory addresses.
For Intel GPUs, the same task can be performed using the \texttt{dispatch} construct from the OpenMP API 5.1.
We note that the two directive-based models share similar syntax,
making translation between them straightforward in practice.
After the GEMV completes, \texttt{update self(y)} (OpenACC) and \texttt{target update from(y)} (OpenMP) transfer the result in \texttt{y} back to the CPU, which is followed by the \texttt{mp\_sum(y,inter\_pool\_comm)} call, a collective summation of \texttt{y}, i.e., $g_{m'n'\ka}(\RRe,\qq)$, among pools.

We note that the \texttt{\_\_CUDA} preprocessor flag in Fig.~\ref{lst:zgemv_gpu} determines at compile time which programming model will be used to manage data movement and offloading.
This porting model, whereby different offloading models are supported via a single source code with preprocessor guards, was successfully employed in the QE suite for the extension of its GPU support to AMD and Intel hardware~\cite{Ruffino2024}.
In the present \epw{} implementation, we find that this approach provides a performance-portable implementation of the first step of the \eph{} matrix interpolation [Eq.~\eqref{eq:g_fine_1}].

\bigskip

\noindent\textbf{Computational settings}

\smallskip

\noindent\emph{Hardware specifications}. Calculations were performed on Perlmutter, Vista, and Aurora supercomputers.
Perlmutter CPU and GPU nodes are equipped with AMD EPYC 7763 (Milan) CPUs and NVIDIA A100 GPUs, while Vista nodes include NVIDIA Grace CPUs and NVIDIA H200 GPUs.
Aurora computing nodes employ Intel Xeon Max 9470C CPUs and Intel Data Center Max 1550 GPUs.
Each system relies on a different software stack: NVIDIA HPC Software Development Kit (SDK) for x86 and Arm servers on Perlmutter and Vista, respectively, and Intel oneAPI HPC Toolkit on Aurora.
Other detailed hardware specifications of these systems are summarized in Tab.~\ref{tab:gpu_node}.

\smallskip

\noindent\emph{Optimal parallelization parameters}. Table~\ref{tab:parallel_params} summarizes our choices of optimal parallelization parameters, i.e., $\Nim$, $\Npo$, and $\Nth$, across \epw{} 5.9, 6.0, and 6.1 on the three benchmark systems.
We further note that a single GPU node in Vista has one H200 GPU (Tab.~\ref{tab:gpu_node}), and thus the one-to-one mapping between MPI rank and GPU restricts the number of images per node to 1 ($\Nim=1$), which can underutilize the peak performance of the GPU.
To circumvent this limitation, on Vista we used NVIDIA's Multi-Process Service (MPS), which allows multiple MPI ranks to share a GPU's resources.
We found that $\Nim=4$ reduced the wall time for the interpolation by $\sim$51~\%,
and we adopted this value for both the \epw{} 6.0 and \epw{} 6.1 benchmarks on Vista.

\smallskip

{
\noindent\textbf{First-principles calculations for zigzag stanene nanoribbons}

\smallskip

First-principles calculations for hydrogen-passivated zigzag stanene nanoribbons were performed using QE~\cite{QE2009,QE2017,QEGPU} within the framework of DFT and DFPT.
Spin-orbit coupling was included through fully relativistic optimized norm-conserving Vanderbilt pseudopotentials~\cite{Hamann2013} with the Perdew-Burke-Ernzerhof (PBE) exchange-correlation functional~\cite{PerdewPBE1996}.
Structural relaxations were performed for all ribbon widths with a vacuum region of 12~\AA{} along the non-periodic directions ($\vec{a}$ and $\vec{c}$ axes).
Self-consistent field calculations employed a $1 \times 12 \times 1$ $\mathbf{k}$-point grid, while phonon calculations were performed on a $1 \times 3 \times 1$ $\mathbf{q}$-point grid.
The coarse-grid parameters for Wannier interpolation, including the number of Wannier bands, phonon modes, and Wigner-Seitz vectors, are summarized in Table~\ref{tab:gsize_zsnr}.
The iterative Boltzmann transport equation was solved within an energy window of 0.8~eV centered at the Fermi level using the adaptive smearing method~\cite{LiShengBTE2014}.
The fine $\mathbf{k}$ and $\mathbf{q}$ grids for Wannier interpolation were set to  $1 \times 500 \times 1$.
}

\bigskip

\noindent\textbf{DATA AVAILABILITY}

\smallskip

\noindent
The raw data for charge carrier transport and electron self-energy benchmark calculations are available from the corresponding author upon reasonable request. All configuration and input files will be publicly available on the Materials Cloud at publication.

\bigskip

\noindent\textbf{CODE AVAILABILITY}

\smallskip

\noindent
The code described and used in this work is publicly available under GNU GPL license as EPW v6.1 (epw-code.org).

\bigskip

\noindent\textbf{REFERENCES}

\bibliography{main} 

\bigskip

\noindent\textbf{ACKNOWLEDGEMENTS}

\smallskip

\noindent
This work was supported by the Computational Materials Sciences Program of the U.S. Department of Energy, Office of Science, Basic Energy Sciences, under Award No. DE-SC0020129. We used resources of the National Energy Research Scientific Computing Center and the Argonne Leadership Computing Facility, which are DOE Office of Science User Facilities supported by the Office of Science of the U.S. DOE, under Contracts DE-AC02-05CH11231 and DE-AC02-06CH11357, respectively; and the Texas Advanced Computing Center at the University of Texas, Austin. We thank Ye Luo, Paolo Giannozzi, Hang Liu, and the EPW Collaboration for helpful discussions on code development and benchmarks.

\bigskip

\noindent\textbf{AUTHOR CONTRIBUTIONS}

\smallskip

\noindent
T.Y.K. designed the hybrid MPI-GPU-OpenMP framework and implemented the GPU and OpenMP accelerations. Z.L. and E.R.M. developed the scheme for distributing the coarse-grid electron–phonon matrix. S.T. implemented the two-level MPI parallelization. T.Y.K. performed the topological stanene nanoribbons calculations. T.Y.K. and S.T. carried out the benchmark simulations. F.G. designed and supervised the project. All authors contributed to writing the manuscript.

\bigskip

\noindent\textbf{COMPETING INTERESTS}

\smallskip

\noindent
The authors declare no competing financial or non-financial interests.

\clearpage
\onecolumngrid
\appendix

\clearpage
\begin{table}[]
    \centering
    \begin{tabular}{l @{\hspace{0.2cm}} l @{\hspace{0.4cm}} l @{\hspace{1.3cm}} l @{\hspace{0.8cm}} l}
        \hline\hline\\[-6pt]
        Node type & Component & Perlmutter (NERSC) & Vista (TACC) & Aurora (ALCF)\\[2pt]
        \hline\\[-6pt]
        CPU+GPU & CPU & AMD EPYC 7763 & NVIDIA Grace C1 & 2$\times$ Intel Xeon Max 9470C \\
        & \hspace{5pt}Cores & 64 & 72 & 104 \\
        & \hspace{5pt}Peak FLOPS & 0.9~TFLOPS & {3.5~TFLOPS} & 5~TFLOPS \\
        & \hspace{5pt}CPU DRAM & 256~GB & 120~GB & 1~TB \\
        & \hspace{5pt}{DRAM BW} & 0.2~TB/s & {0.5~TB/s} & 0.56~TB/s \\
        & \hspace{5pt}CPU HBM & - & - & 128~GB \\
        & \hspace{5pt}HBM BW & - & - & 2.87~TB/s \\[4pt]
        & GPU & 4$\times$ NVIDIA A100 & NVIDIA H200 & 6$\times$ Intel Data Center Max 1550 \\
        & \hspace{5pt}Peak FLOPS & 40~TFLOPS & 34~TFLOPS & 302~TFLOPS \\
        & \hspace{5pt}GPU HBM & 160~GB & 96~GB & 768~GB \\
        & \hspace{5pt}HBM BW & 6.4~TB/s & {4~TB/s} & 19.66~TB/s \\[2pt]
        \hline\\[-6pt]
        CPU only & CPU & 2$\times$AMD EPYC 7763 & NVIDIA Grace Superchip \\
        & \hspace{5pt}Cores & 128 & 144 \\
        & \hspace{5pt}Peak FLOPS & 1.8~TFLOPS & {7~TFLOPS} \\
        & \hspace{5pt}CPU DRAM & 512~GB & 240~GB \\
        & \hspace{5pt}DRAM BW & 0.4~TB/s & {1~TB/s} \\[2pt]
        \hline\hline
    \end{tabular}
    \caption{\textbf{Hardware specifications of single GPU-accelerated and CPU-only computing nodes of the supercomputers used in the present benchmarks.} Perlmutter is at the U.S. National Energy Research Scientific Computing Center (NERSC), Vista is at the Texas Advanced Computing Center (TACC), and Aurora is at the Argonne Leadership Computing Facility (ALCF).  Note that Aurora at the Argonne Leadership Computing Facility (ALCF) does not have a dedicated CPU-only node.}
    \label{tab:gpu_node}
\end{table}

\clearpage
\begin{table}[htbp]
    \centering
    \begin{tabular}{l @{\hspace{0.7cm}} ccc @{\hspace{0.7cm}} ccc @{\hspace{0.7cm}} ccc}
        \hline\hline\\[-6pt]
        & \multicolumn{3}{c}{Perlmutter (NERSC)} 
        & \multicolumn{3}{c}{Vista (TACC)} 
        & \multicolumn{3}{c}{Aurora (ALCF)} \\
        EPW version & $\Nim$ & $\Npo$ & $\Nth$ & $\Nim$ & $\Npo$ & $\Nth$ & $\Nim$ & $\Npo$ & $\Nth$ \\[2pt]
        \hline\\[-6pt]
        5.9 (CPU-only) & – & 128 & – & – & 144 & – & – & 96 & – \\
        6.0 (CPU-only) & 4 & 32  & – & 16 & 9   & – & 12 & 8  & – \\
        6.1 (CPU+GPU) & 4 & 1   & 16& 4  & 1   & 18& 12 & 1  & 8 \\[2pt]
        \hline\hline
    \end{tabular}
    \caption{\textbf{Optimal parallelization parameters in single-node benchmarks.} 
    Benchmarks were performed on the Perlmutter, Vista, and Aurora supercomputers using three different versions of EPW.
    $\Nim$ is the number of images, $\Npo$ the number of pools, and $\Nth$ the number of OpenMP threads per rank. 
    The listed parameters yielded optimal performance for each combination of EPW version and system.}
    \label{tab:parallel_params}
\end{table}

\clearpage
\begin{table}
    \centering
    \begin{tabular}{c @{\hspace{0.8cm}} r @{\hspace{0.6cm}} r @{\hspace{0.6cm}} r @{\hspace{0.6cm}} r @{\hspace{0.8cm}} c}
        \hline\hline
        Width (nm) & $\Nb$ & $\Nm$ & $\NRRe$ & $\NRRp$ & $g_{m'n'\ka}(\RRe,\RRp)$ [GB] \\
        \hline
        3.3  & 32 & 54 & 57 & 15 & 0.7 \\
        4.9  & 48 & 78 & 57 & 15 & 2.3 \\
        6.5  & 64 & 102 & 59 & 15 & 5.5 \\
        8.1  & 80 & 126 & 79 & 15 & 14 \\
        9.7  & 96 & 150 & 77 & 15 & 24 \\
        11.4 & 112 & 174 & 97 & 15 & 47 \\
        13.0 & 128 & 198 & 135 & 15 & 98 \\
        14.6 & 144 & 222 & 153 & 15 & 157 \\
        16.2 & 160 & 246 & 169 & 15 & 238 \\
        17.8 & 176 & 270 & 189 & 15 & 353 \\
        19.4 & 192 & 294 & 189 & 15 & 458 \\
        \hline\hline
    \end{tabular}
    \caption{
        {\textbf{Size of coarse-grid electron-phonon matrix used in phonon-limited carrier transport calculations for hydrogen-passivated zigzag stanene nanoribbons of various widths.}
        The width represents the distance between two hydrogen atoms at the opposite zigzag edges in the unit cell.
        $\Nb$, $\Nm$, $\NRRe$, $\NRRp$ are the number of Wannier bands, phonon modes, and electron and phonon Wigner-Seitz vectors, respectively.
        The size of $g_{m'n'\ka}(\RRe,\RRp)$ is given by $16 \, \Nb^2 \Nm \NRRe \NRRp$ bytes.
        }
    }
    \label{tab:gsize_zsnr}
\end{table}

\clearpage

\clearpage

\clearpage

\clearpage

\clearpage

\clearpage

\clearpage

\clearpage

\clearpage
\begin{figure}
    \centering
    \includegraphics{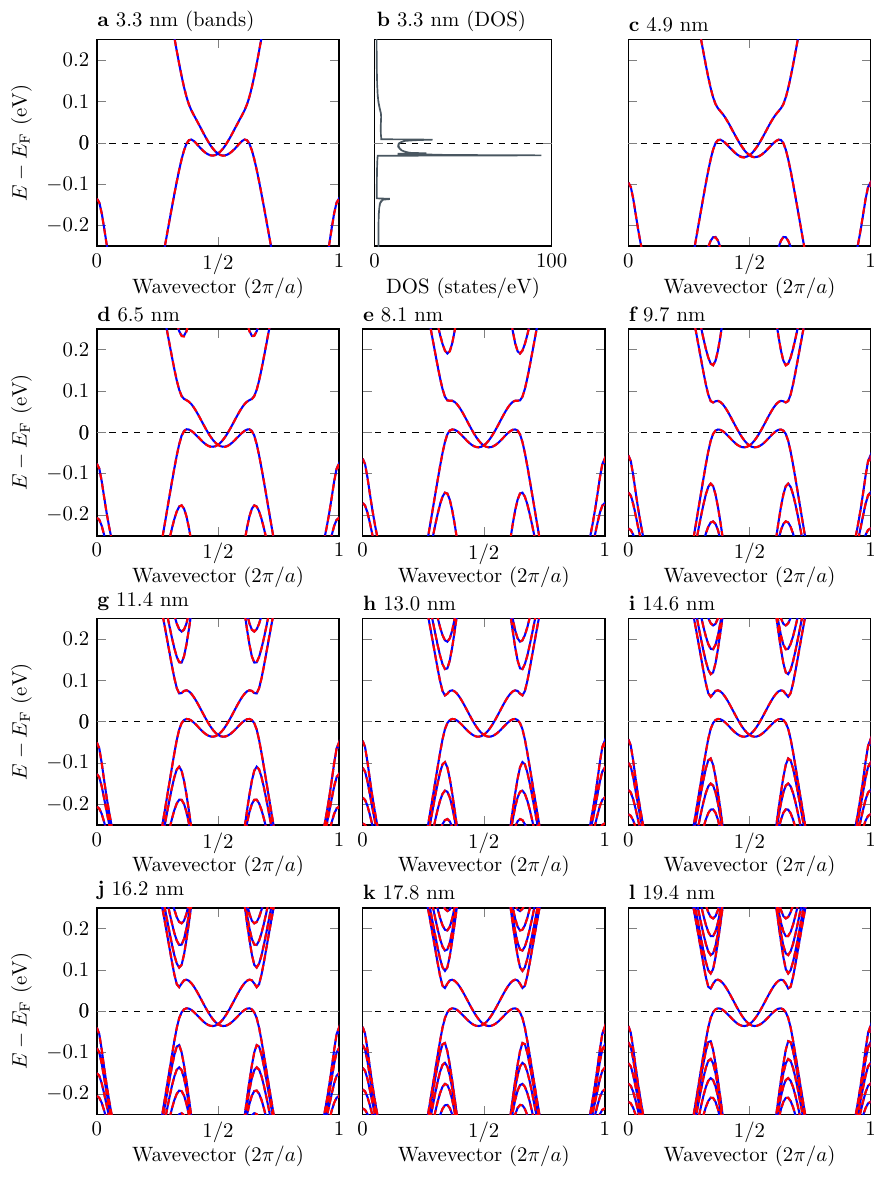}
    \caption{
        \textbf{Band structure and density of states of hydrogen-passivated zigzag stanene nanoribbons.}
        {\bf a},\,{\bf c}--{\bf l} Comparison between density functional theory (blue solid) and Wannier-interpolated (red dashed) band structures for zigzag stanene nanoribbons of various widths from 3.3~nm to 19.4~nm.
        {\bf b} The density of states for the 3.3~nm ribbon.
        The width values represent the separation of the two hydrogen atoms at opposite zigzag edges.
        The horizontal axes for the band structures are in units of $2\pi/a$, where $a$ is the unitcell length along the periodic direction (the $\vec{b}$ axis).
    }
    \label{fig:band_comparison}
\end{figure}

\clearpage
\begin{figure}
    \centering
    \includegraphics{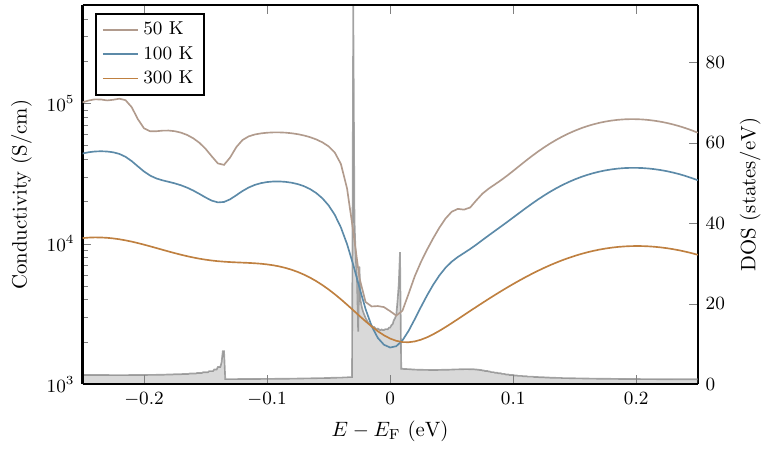}
    \caption{
        {\textbf{Calculated phonon-limited conductivity and density of states (DOS) of hydrogen-passivated zigzag stanene nanoribbon of width 3.3~nm.}
        The conductivity (left axis) at 50~K, 100~K, and 300~K, together with the DOS (shaded, right axis) are shown as a function of energy relative to the Fermi level.}
    }
    \label{fig:zsnr_dos}
\end{figure}

\clearpage
\begin{figure}
    \centering
    \includegraphics{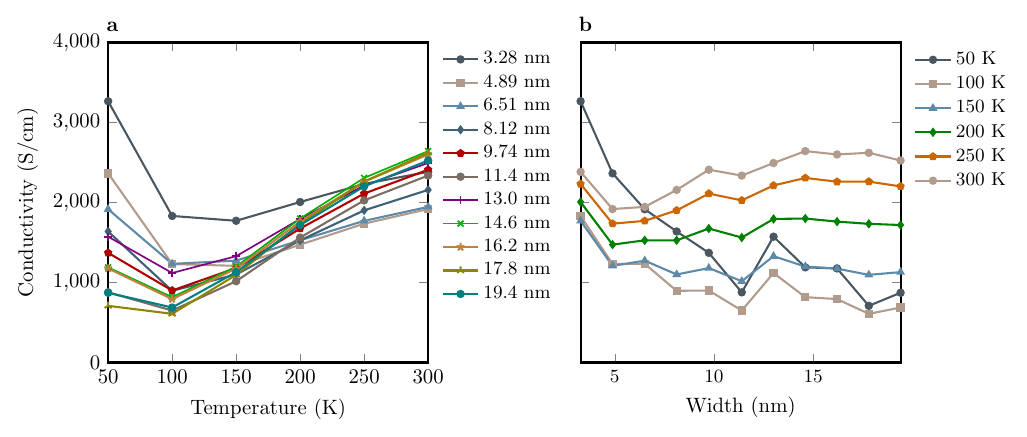}
    \caption{
        \textbf{Phonon-limited conductivity of hydrogen-passivated zigzag stanene nanoribbons.}
        {\bf a} Conductivity as a function of ribbon width for temperatures of 50, 100, 150, 200, 250, and 300 K.
        {\bf b} Conductivity as a function of temperature for ribbon widths ranging from 3.3 to 19.4 nm.
    }
\label{fig:sigma_combined}
\end{figure}

\clearpage

\clearpage

\end{document}